\documentclass{aastex631}

\usepackage{graphicx} % Required for inserting images
\usepackage{amsmath}
\usepackage{xcolor}
\usepackage{relsize}
\usepackage{xspace}

\newcommand{\todo}[1]{}
\newcommand{\bfedit}[1]{#1}

\def\CTA{Center for Theoretical Astrophysics, Los Alamos National Laboratory, Los Alamos, NM 87545, USA}
\def\CCS{Computer, Computational, and Statistical Sciences Division, Los Alamos National Laboratory, Los Alamos, NM
 87545, USA}
\def\CNLS{Center for Nonlinear Studies, Los Alamos National Laboratory, Los Alamos, NM
 87545, USA}
\def\T5{T-5 Applied Mathematics and Plasma Physics Group, Los Alamos National Laboratory, Los Alamos, NM 87545, USA}
\def\XCP{Computational Physics Division, Los Alamos National Laboratory, Los Alamos, NM, 87545, USA}
\def\UA{The University of Arizona, Tucson, AZ 85721, USA}
\def\NM{Department of Physics and Astronomy, The University of New Mexico, Albuquerque, NM 87131, USA}
\def\GWU{The George Washington University, Washington, DC 20052, USA}
\def\TSI{Trottier Space Institute at McGill, McGill University, 3550 rue University, Montreal, Québec, H3A 2A7, Canada}

\def\CIERA{Center for Interdisciplinary Exploration and Research in Astrophysics, Northwestern University, 1800 Sherman Ave., Evanston, IL 60201, USA}
\def\LSU{Department of Physics and Astronomy, Louisiana State University, Baton Rouge, LA 70803, USA}

\def\eg{{\it e.g.}}

\def\ie{{\it i.e.}}

\begin{document}

\title{A comparison of three neodymium atomic data sets for kilonova modeling}

\author[0000-0003-1087-2964]{Christopher~J. Fontes}
\affiliation{\CTA}
\affiliation{\XCP}

\author[0000-0001-7815-7604]{Nicholas Vieira}
\affiliation{\TSI}
% \affiliation{\MCGILL}
\affiliation{\CIERA}

\author[0000-0003-2624-0056]{Chris~L. Fryer}
\affiliation{\CNLS}
\affiliation{\CTA}
\affiliation{\UA}
\affiliation{\NM}
\affiliation{\GWU}

\author[0000-0002-8560-692X]{Adithan Kathirgamaraju}
\affiliation{\CTA}
\affiliation{\CCS}

\author[0000-0003-4156-5342]{Oleg Korobkin}
\affiliation{\CTA}
\affiliation{\T5}

\author[0000-0001-7042-4472]{Marko Ristic}
\affiliation{\CTA}
\affiliation{\T5}

\author[0000-0003-3265-4079]{Ryan~T. Wollaeger}
\affiliation{\CTA}
\affiliation{\CCS}
\affiliation{\LSU}

\begin{abstract}
    We examine the impact of input neodymium (Nd) atomic data on the light curves and spectra of kilonovae, probing the sensitivity of kilonova observables to the atomic physics of this important lanthanide element. We use the {\tt SuperNu} Monte Carlo radiative transfer code, simulating a simple semi-analytic 1D kilonova with a pure Nd atmosphere, fixing the radiative
    transfer method while using input atomic data generated by three different codes: the LANL suite of atomic physics codes, \texttt{HULLAC}, and \texttt{Autostructure}. We see that the choice of atomic data significantly shapes the resulting light curves and spectra. Peak bolometric luminosities differ by a ratio of nearly 1.5 between \texttt{HULLAC}/\texttt{Autostructure} and LANL data sets. Moreover, we observe significant near- to mid-IR differences in the structure of the spectra. We specifically attribute these differences to the choice of atomic data for neutral Nd~I. Many of the results here have been adapted from a presentation at ``Radiative Transfer and Atomic Physics of Kilonovae'' in Stockholm, 2023. We additionally present a LANL data set with energies calibrated to available values in the NIST Atomic Spectra Database, and demonstrate that this calibration also significantly affects IR spectral structure at late time. The substantial differences in kilonova observables that arise from tuning the atomic data of just one lanthanide element highlight the special attention that must be paid to atomic physics uncertainties when modeling kilonovae, from AT2017gfo to beyond.
    %\todo{summarize new spectroscopically calibrated line data.}
\end{abstract}

\section{Introduction}

%-- general kilonova intro
Kilonovae (KNe) are the UV/optical/IR transients produced during binary neutron star and neutron star black hole mergers
\citep{lattimer1974}. Such a kilonova was observed in conjunction with a gravitational wave signal for the first time in the August 2017 event GW170817 (see, \eg, 
\citealt{arcavi2017,cowperthwaite2017,drout2017,kasliwal2017,smartt2017,tanvir2017,troja2017,villar2017}).
These transients are powered by the thermalization of the radioactive decay products of nuclei produced by rapid neutron capture
($r$-process) nucleosynthesis (\eg, \citealt{freiburghaus1999}).
This $r$-process produces a suite of heavy-element isotopes with a range of radioactive decay time
scales such that the rate of kilonova ejecta heating from decay products is reasonably described by a simple temporal power law
(\eg, \citealt{metzger2010,korobkin2012}, but see \citealt{rosswog2024}). Hence a simple semi-analytic model for the bolometric luminosity may be formulated as well, as was
initially done by~\cite{li1998}.

%-- literature review (use recent review slides to start)
%\todo{Re-read some of this...}
Recent effort in modeling KNe has focused on using spectroscopically accurate, experimentally validated
atomic data to generate synthetic spectra or light curves for comparison to observation.
An important early step in the spectroscopic characterization of AT2017gfo, the electromagnetic counterpart to GW170817, was carried out in ~\cite{watson2019}. They
show that a strontium II atomic bound-bound transition (line) can yield a P Cygni feature observed in the spectra of AT2017gfo at $\sim$0.7-1 \textmu m. Similarly, at $\sim$2.1 \textmu m, \cite{hotokezaka2023} identify a strong emission feature that may be reproduced by with tellurium (Te) III at late times. Te III is also identified in the late-time spectra of the KN associated with
gamma-ray burst GRB 230307A (\citealt{gillanders2023}). \cite{gillanders2023} also find features with candidates including
Te II at 4.4 \textmu m and cerium III, thorium (Th) III, or platinum II at 2.1 \textmu m.
Recently,~\cite{domoto2025} used experimental line data to identify a Th III absorption feature at
$\sim$1.8 \textmu m as a candidate for future detection with the James Webb Space Telescope (JWST).

A common approach among these recent works in spectroscopic characterization of KNe is to use experimental opacity data, theoretical data that have been calibrated to experiment, or some combination of both. Given the incompleteness of experimental atomic data, especially for the elements of relevance in KN ejecta, theoretical calculations are of particular value. Indeed, the work of~\cite{domoto2025} recently prompted theoretical opacity calculations
for Th~III~\citep{kitoviene2025}.
Significant further effort in the relevant atomic physics calculations is pursued by
\cite{deprince2025}, who find higher Planck mean opacities than previous groups (\eg, \citealt{tanaka2020, fontes2020}), but energy levels within 10--20\% of those computed with the Flexible Atomic Code (FAC;~\citealt{gu2008}), for several ions of the lanthanides---elements of atomic number $Z$~from 57 to 71, with substantial contributions to KN opacities anticipated.
\texttt{FAC} has recently been used to compute new dielectronic recombination (DR) rates, which are relevant to correcting the ionization balance in late-time, nebular KNe~\citep{singh2025}.

%-- motivation
The lanthanide neodymium (Nd; $Z = 60$) is a potentially significant contributor to absorption and emission in KNe~\citep{even2020,deprince2025}.
In particular, it may contribute to the ``lanthanide curtaining'' effect~\citep{kasen2015}, in which optical emission (\eg, post-merger accretion disk wind with higher electron fraction) is blocked by a curtain of faster, lower-electron fraction, redder material (\eg, tidally ejected material) and reprocessed into IR emission. Though most studies of Nd explore the element in the KN thermodynamic regime, it has also has been the focus of studies out of local thermodynamic equilibrium, \ie, in non-LTE~\citep{hotokezaka2021}. Calculations in and out of LTE benefit from the existence of some experimental data for non-neutral ion stages of Nd (\eg,~\citealt{flors2023,maison2024}).

The potentially important role of Nd in KN physics motivates a comparison of bound-bound (line) atomic data for the element from the various different theoretical data sets that have been used in the KN literature. This sort of comparison, along with observational impact, has been done for multi-element KN models (\citealt{brethauer2024}). In the present study, adapted in part from a presentation at the ``Radiative Transfer and Atomic Physics of Kilonovae'' workshop in Stockholm in 2023, we focus solely on Nd. Using radiative transfer simulations with the {\tt SuperNu} code, we explore the impact of Nd on kilonova observables for three different input atomic data sets: a set generated with LANL codes, another with the \texttt{HULLAC} code, and another with the \texttt{Autostructure} code. We observe the strong impact of Nd on our KN light curves and spectra, especially at later times ($\sim$$1$~week post-merger), and how this impact varies for different input atomic data.
In particular, we see strong line features in the near- to mid-IR range in our models~\citep{korobkin2021}
and demonstrate that these features correspond to bound-bound transitions in Nd, and especially its neutral stage (Nd I). Using a simple, 1D KN model with a pure Nd atmosphere in {\tt SuperNu} permits us to isolate the element, while also not
over-weighting atomic differences that may not contribute significantly to emission.

%-- navigation paragraph
This article is organized as follows.
In Section~\ref{sec:methods}, we provide a brief overview of the three atomic physics codes that generate the data used in this study and describe our techniques for calibrating our LANL calulated energies to curated NIST data. We also briefly summarize the {\tt SuperNu} radiative transfer code. In Section~\ref{sec:numres}, we compare basic aggregate statistics among data sets and use a simple
1D KN model with a pure Nd atmosphere to examine how the atomic data affect ejecta thermodynamics, light
curves, and spectra. We also isolate the impact of neutral Nd and examine the effect of the calibration of LANL-calculated energies to NIST data.
Finally, in Section~\ref{sec:conc} we summarize and conclude.

\section{Methods}
\label{sec:methods}

\subsection{Atomic Nd line data}
\label{sec:nddata}

We compare Nd data sets that contain the first four ion stages, \ie, the neutral stage through triply ionized. The data sets were generated with three different computational frameworks that have been used to generate atomic data over several decades: 
the LANL suite of atomic physics codes~\citep{fontes2015,fontes2015b}, the \texttt{HULLAC} code~\citep{barshalom2001} and the \texttt{Autostructure} code~\citep{badnell2011,badnell2012,badnell2016}.
The codes and data set sources are briefly summarized here and in the following subsections; further elaboration
can be found in the references.

The basic approach for calculating the atomic structure with each of these three methods follows the well-known configuration-interaction (CI) theory (\eg,~\citealt{cowan1981,grant2007,sampson2009,fontes2015b}). A structure calculation begins with the specification of a list of relevant atomic configurations for each ion stage. Each configuration is specified by a set of electron orbitals and their integer occupation numbers. The orbitals can be specified using non-relativistic ($nl$) or relativistic ($nlj$) orbital notation, where the former are used when solving the Schr\"odinger equation and the latter when solving the Dirac equation. Specifically, radial wavefunctions for each orbital in a configuration are obtained by solving the appropriate equation, which contains an electron-electron potential that represents the interaction between all of the electrons in a configuration. Some codes obtain a single set of orbital radial wavefunctions to represent the entire list of specified configurations, while other codes obtain a different set of orbital radial wavefunctions for each configuration in the list.

These electron orbitals are then combined, via standard angular-momentum coupling techniques, to produce single-configuration state functions (SCSFs). The coupling is typically done using $LS$ coupling for $nl$-type configurations and $jj$ coupling for $nlj$-type configurations. The result of this procedure is a set of pure, so-called basis states, which are characterized by their parity and $J$ value, the total angular momentum quantum number.
The complete $N$-electron Hamiltonian is then expressed numerically in a subspace of the SCSFs for all basis states with a given parity and $J$ value. Approximate, fine-structure
ion wave functions are obtained by diagonalizing this Hamiltonian. The result of this diagonalization process is a set of fine-structure energies and wavefunctions, which are represented as linear combinations of the SCSF basis states. The mixing of pure basis states that arise from different configurations is a hallmark of the CI approach.
Finally, these fine-structure wavefunctions can be used to evaluate dipole matrix elements to obtain the oscillator strengths that are necessary to calculate the bound-bound (or line) contribution to the opacity.

A summary of atomic model features relevant to the data sets used in this study is given in
Table~\ref{tab1:nddata}.
We note that these model properties do not constitute an exhaustive list of the potential contributors
of differences observed among the data sets; these properties are merely some items to consider in the
comparison of this data, which are actively used in various KN modeling efforts.
\begin{table}[]
    \centering
    \begin{tabular}{|c|c|c|c|}
        \hline
        Code/data & Hamiltonian type & Single-electron effective potential & Angular momentum coupling \\
        \hline
        LANL & Semi-relativistic Schr\"odinger & Hartree–Fock & Racah algebra ($LS$-coupling) \\
        \texttt{HULLAC} & Dirac & Parametric (NIST calibrated) & Racah algebra ($jj$-coupling) \\
        \texttt{Autostructure} & Breit–Pauli & Thomas–Fermi–Dirac–Amaldi & Slater state representation \\
        \hline
    \end{tabular}
    \caption{Code/data (row) and atomic model information (column) compared in the present study.}
    \label{tab1:nddata}
\end{table}

%\todo{In following subsections, maybe add the configuration lists (at least for the Autostructure data, which is unpublished as far as I can tell).}

\subsubsection{LANL suite of atomic physics codes}
\label{sec:lanlsapc}

The LANL suite of atomic physics codes~\citep{fontes2015b} contains both semi-relativistic~\citep{cowan1981} and fully relativistic~\citep{sampson2009} capabilities to calculate atomic structure. These are ab initio calculations that do not include any tuning or calibration to experimental or curated energies.
This suite has been used to calculate opacities under KN ejecta
conditions, in particular for a set of lanthanide~\citep{fontes2020} and actinide
\citep{fontes2023} opacity tables over density, temperature and frequency.\footnote{\bfedit{See \cite{ralchenko2020}.}}

In the present study, the {\tt CATS} (Cowan's ATomic Structure) code~\citep{cats89}, which employs the semi-relativistic Hartree-Fock (HFR) approach of \cite{cowan1981},
was used to generate the Nd line list. The relevant list of Nd configurations was specified by~\cite{fontes2020}, who duplicated the list chosen by \cite{kasen2013} in order to provide meaningful comparisons with the latter work.
We also attempt to improve the ab initio HFR model by using calibrated energies, \ie, the calculated HFR level energies are replaced with available energies from the NIST Atomic Spectra Database (ASD;~\citealt{nist_asd}).
Our calibration approach is described in Section~\ref{sec:callanl}.

\subsubsection{\textsc{HULLAC}}

\texttt{HULLAC} (Hebrew University Lawrence Livermore Atomic Code) is described by~\cite{barshalom2001}, and employs a parametric potential in the radial-wavefunction equations that are obtained from the Dirac equation.
%non-perturbative radial potential expressed as a parameterized sum over closed and open orbital shells.
%Since these shell parameters set the radial wavefunctions, calculations can be optimized by an energy minimization
%procedure, for instance minimizing differences with experimental data~\citep{barshalom2001}.
The parameters associated with each $nlj$ orbital of interest can be optimized by an energy minimization procedure, for instance minimizing differences with experimental data~\citep{barshalom2001}.
%At the single-electron wavefunction level, the relativistic Dirac equation is solved with this effective potential.

\cite{tanaka2020}\footnote{\bfedit{See~\cite{kato2021}.}} applied the \texttt{HULLAC}
code to generate a suite of opacities from proton number $Z=26$ to 88, and compared the ionization
energies with those provided in the NIST ASD~\citep{nist_asd}.
\bfedit{Recently~\cite{banerjee2025} applied \texttt{HULLAC} to computing DR rates for nebular non-LTE modeling.}
While the Japan-Lithuania database also includes corresponding atomic data generated with the {\tt GRASP2K} code~\citep{gaigalas2019}, for the present study, we use the \texttt{HULLAC} data set because it includes a line list for the neutral stage of Nd.
\bfedit{We note that~\cite{kato2024} have recently explored improvements to the \texttt{HULLAC} calculations, altering the effective potential and number of configurations, in the context of singly ionized lanthanides. We do not use these improved HULLAC data here, but note that the opacities resulting from these improved data can be higher by a factor of 3--10 (see their Figure 4 in particular for Nd II), which would affect the comparisons in Section~\ref{sec:numres} near peak bolometric luminosity (occurring around days~2 to 5).}

\subsubsection{\textsc{Autostructure}}

\texttt{Autostructure}~\citep{badnell2011,badnell2012,badnell2016}, based on the \texttt{SUPERSTRUCTURE} code~\citep{eissner1974},
is similar in capability to the prior two codes. It also permits specification of different types of effective 
potentials for computing the electron-orbital radial wavefunctions.
For the orbital radial wavefunctions, it can employ either non-relativistic or ``kappa-averaged''~\citep{cowan1981} relativistic
formulae.
%In~\cite{badnell2011}, it was noted that Autostructure is efficient in particular for computing dielectric recombination
%(see also~\cite{badnell1986}).
\cite{badnell2012} improved the efficiency for open $f$-shell computation by using a Slater state approach to
angular momentum coupling (as apposed to hierarchical Racah algebra) and permitting certain memory-saving
optimizations (for instance, combining states) during the atomic calculation.

The ability to perform efficient calculations for elements with open $f$-shell orbitals
\citep{badnell2012} has enabled \texttt{Autostructure} to be used in a multitude of KN studies (see, for instance, the first such uses in \citealt{kasen2013,barnes2013}). 
\cite{kasen2013} used the Thomas–Fermi–Dirac–Amaldi effective potential, which includes electron correlation
and exchange effects (\citealt{eissner1974}; see, for instance,~\citealt{bautista2008}).

In the calculations used here, we use the kappa-averaged relativistic wavefunctions and the Thomas–Fermi–Dirac–Amaldi effective potential.
For optimization, we emulate the scheme outlined in \cite{kasen2013} as ``opt1'', whereby we simultaneously \bfedit{adjust} the scaling parameters for all included orbitals during optimization. Furthermore, we employ the same configuration list as~\cite{tanaka2020}.
For completeness, we note that we attempted a preliminary study with the \texttt{Autostructure} data used by \cite{kasen2013}, but that data set was incomplete for the purposes of this study.

\subsection{Calibration of LANL energies}
\label{sec:callanl}

Since it is typically difficult for atomic structure codes to accurately predict, in an ab initio manner, the energies and ordering of the levels of lowly charged lanthanide and actinide elements, it can be beneficial to incorporate more accurate energies in some sort of calibration technique. The use of more accurate energies provides the added benefit of improving the oscillator strengths, which depend linearly on the transition energy between the associated pair of levels. In the present work, our calibration technique involves the replacement of ab initio energies with more accurate energies. While it is impossible to replace all of the calculated energies, it is desirable to improve, at the very least, the energies of the lowest lying levels of a given charge state. Transitions among these levels often provide spectral lines that are useful for elemental identification and diagnosing plasma conditions.

The energy-calibration technique employed here can be rather challenging for the low charge states of lanthanide elements, such as Nd, due to the difficulty in mapping the NIST ASD energies to the corresponding energies in the {\tt CATS} calculations. The energy levels in the NIST ASD are typically labeled according to standard $LS$-coupling angular momentum notation. Each level is identified by a dominant electron configuration, an $LS$ term, and a total angular momentum quantum number, $J$.
If only a single configuration is considered in an atomic structure calculation, then all of the fine-structure levels associated with a given $LS$ term are referred to as a multiplet, which is a useful concept because the NIST ASD groups energies in this way.
For example, according to the NIST ASD, the lowest energy multiplet for Nd~I arises from the $4f^4\, 6s^2$ configuration and has an $LS$ term label of $^5I$. There are five levels denoted by integer $J$ values ranging from 4--8.

However, the situation is complicated when the aforementioned CI approach is used to calculate fine-structure energy levels. If the mixing of basis states is relatively pure, such that an unambiguous labels can be obtained, then the concept of a multiplet is still valid and a clean mapping can be made between NIST and {\tt CATS} energies.
On the other hand, if the mixing between configurations is so strong that the linear combinations of basis states that represent various levels do not produce clear, dominant labels, then the concept of multiplets breaks down and it can be very difficult to map NIST energies to {\tt CATS} energies.
In that case, it is not always possible to cleanly assign a dominant configuration and $LS$ term to each calculated level.
The linear combination of basis states associated with a given energy level can be so complicated that there is no obvious choice of a dominant label.
In such cases, one must make an educated guess as to which levels reside in a given multiplet, based on information such as the relative energy spacings that would have occurred among the levels in a multiplet if a single-configuration calculation (see earlier discussion) had been carried out. 

The mapping process is further complicated by the fact that the same $LS$ term can arise more than once from a given configuration, in which case additional, intermediate $LS$ terms are typically required to uniquely label the energy levels in the different multiplets. Such intermediate terms are not always easy to map from {\tt CATS} to the ASD, if they exist there. Another challenge arises from the fact that the NIST ASD does not provide a complete set of energy levels. For example, sometimes a multiplet that is listed in the ASD does not contain energies for all of the $J$-resolved levels in a multiplet and sometimes there are entire multiplets that are missing from the ASD.
The former case can be addressed by filling in the missing energies of a multiplet by assuming that the relative energy spacing in the NIST data should be the same as the relative energy spacing in the {\tt CATS} calculations. The latter case is more problematic and leaves open the possibility that some of the calculated {\tt CATS} energies will be shifted via the calibration procedure, while others remain unchanged.
One remedy is to simply shift the energies of all of the levels that reside in a missing multiplet that has the same $LS$ term and dominant configuration as a multiplet that does exist in the ASD, by assuming that the relative energy difference between, say, the lowest energy in each multiplet should be preserved. Of course, this remedy requires at least one $LS$ term in the list of duplicate $LS$ terms to exist in the ASD.

The above calibration options were particularly challenging to apply to Nd~I and II, but, to our knowledge, they offer a first attempt to obtain a set of explicitly calibrated low-lying levels, which are typically the most important for interpreting KN spectra.
We also note that other authors \citep{kasen2013,tanaka2018,deprince2023,flors2023} have investigated energy-calibration techniques that are somewhat more general in nature, which has the appeal of self-consistency, rather than the present approach which attempts to map specific NIST energies to specific calculated energies.
For example, the energy-calibration study of Nd~II and III by \cite{flors2023} included level energies that were calculated with the HFR method of Cowan that is also used in the present study. However, that early study used a calibration technique based on the work of \cite{deprince2023} which matches the calculated configuration-average energies with those deduced from experimental level energies.

\bfedit{Using the above procedures, the following number of levels were calibrated for each of the four charge states considered in this work: 217 levels for Nd~I, 187 levels for Nd~II, 26 levels for Nd~III and 18 levels for Nd~IV.
These levels ranged in energy from 0.0--3.9808260~eV for Nd~I, 0.0--5.64171~eV for Nd~II, 0.0--4.07072~eV for Nd~III and 0.0--2.657~eV for Nd~IV.

In order to illustrate the degree of difference between the {\tt CATS} and NIST ASD energies, we provide in Tables~\ref{tab:energy_nd1}--\ref{tab:energy_nd4} the first 10 levels, in order of increasing energy, for Nd~I--IV, respectively. As expected, the agreement is worse for the lowest two charge states, and improves for the highest two charge states. For example, the {\tt CATS} code does not predict the NIST ground level for Nd~I and II, but it does for Nd~III and IV. Looking in more detail, for Nd~I there are six levels that are common to the {\tt CATS} and NIST lists displayed in Table~\ref{tab:energy_nd1}, but they appear in different energy order. For example, when the calibration procedure is performed, level number~3 in the LANL {\tt CATS} list $(4f^4 6s^2 \ ^5I_4)$ is remapped to level number~1, i.e. the ground level, in the NIST list.
This mismatch in level ordering makes it difficult to provide a meaningful, quantitative comparison between the energies in the two datasets.
At the other extreme, for Nd~IV all 10 of the levels are present in both the LANL {\tt CATS} and NIST lists. The {\tt CATS} energies are consistently higher than the NIST energies, with differences ranging from 2.6--13\%. Furthermore, the level labels are in almost the same energy order in both lists, except that levels~6 and 7 must be switched, as well as levels~8 and 9, in order to make the {\tt CATS} labels appear in the same order as the NIST labels.}

\begin{table}[h]
\begin{minipage}{6.5in}
\begin{tabular}{|r|r|r|c|c|}
\hline
Level &  LANL  & NIST  & & \\
number & Label & Label & LANL (eV) & NIST (eV) \\
\hline
 1 & $4f^3 5d^1 6s^2 \ ^5L_6$ & $4f^4 6s^2 \ ^5I_4$ & 0.000   & 0.000 \\
 2 &                  $^5K_5$ &             $^5I_5$ & 0.00934 & 0.140 \\
 3 & $4f^4 6s^2 \ ^5I_4$      &             $^5I_6$ & 0.215   & 0.293 \\
 4 & $4f^3 5d^1 6s^2 \ ^5K_6$ &             $^5I_7$ & 0.218   & 0.456 \\
 5 &                  $^5L_7$ &             $^5I_8$ & 0.226   & 0.626 \\
 6 & $4f^4 6s^2 \ ^5I_5$      & $4f^3 5d^1 6s^2 \ ^5L_6$ & 0.368 & 0.839 \\
 7 & $4f^3 5d^1 6s^2 \ ^5I_4$ &                  $^5K_5$ & 0.380 & 0.850 \\
 8 &                  $^5K_7$ &                  $^5L_7$ & 0.433 & 1.042 \\
 9 &                  $^5L_8$ &                  $^5K_6$ & 0.467 & 1.043 \\
10 &                  $^5I_5$ & $4f^4 5d^1 6s^1 \ ^7L_5$ & 0.506 & 1.051 \\
\hline
\end{tabular}
\caption{\bfedit{A list of the first 10 energy levels of Nd~I obtained from
the LANL {\tt CATS} code and the NIST Atomic Spectra Database.}}
\label{tab:energy_nd1}
\end{minipage}
\end{table}

\begin{table}[h]
\begin{minipage}{6.5in}
\begin{tabular}{|r|r|r|c|c|}
\hline
Level &  LANL  & NIST  & & \\
number & Label & Label & LANL (eV) & NIST (eV) \\
\hline
 1 & $4f^3 5d^2 \ ^6M_{13/2}$      & $4f^4 6s^1 \ ^6I_{7/2}$  & 0.000   & 0.000 \\
 2 &             $^6M_{15/2}$      &             $^6I_{9/2}$  & 0.190   & 0.0636 \\
 3 & $4f^3 5d^1 6s^1 \ ^6L_{11/2}$ &             $^6I_{11/2}$ & 0.281   & 0.182 \\
 4 &                  $^6K_{9/2}$  &             $^4I_{9/2}$  & 0.303   & 0.205 \\
 5 &                  $^6L_{13/2}$ &             $^6I_{13/2}$ & 0.385   & 0.321 \\
 6 & $4f^3 5d^2 \ ^6M_{17/2}$      &             $^4I_{11/2}$ & 0.391 & 0.380 \\
 7 & $4f^3 5d^1 6s^1 \ ^6K_{11/2}$ &             $^6I_{15/2}$ & 0.393 & 0.471 \\
 8 & $4f^3 5d^2 \ ^6L_{11/2}$      & $4f^4 5d^1 \ ^6L_{11/2}$ & 0.442 & 0.550 \\
 9 &             $^6K_{9/2}$       & $4f^4 6s^1 \ ^4I_{13/2}$ & 0.476 & 0.559 \\
10 & $4f^4 6s^1 \ ^6I_{7/2}$       &             $^6I_{17/2}$ & 0.496 & 0.631 \\
\hline
\end{tabular}
\caption{\bfedit{A list of the first 10 energy levels of Nd~II obtained from
the LANL {\tt CATS} code and the NIST Atomic Spectra Database.}}
\label{tab:energy_nd2}
\end{minipage}
\end{table}

\begin{table}
\begin{minipage}{6.5in}
\begin{tabular}{|r|r|r|c|c|}
\hline
Level  & LANL  & NIST  & & \\
number & Label & Label & LANL (eV) & NIST (eV) \\
\hline
 1 & $4f^4 \ ^5I_{4}$ &  $4f^4 \ ^5I_{4}$ &0.000 & 0.000 \\
 2 &        $^5I_{5}$ &        $^5I_{5}$ & 0.155 & 0.141 \\
 3 &        $^5I_{6}$ &        $^5I_{6}$ & 0.324 & 0.296 \\
 4 & $4f^3 5d^1 \ ^5L_{6}$  &  $^5I_{7}$ & 0.444 & 0.461 \\
 5 &              $^5K_{5}$ &  $^5I_{8}$ & 0.474 & 0.631 \\
 6 & $4f^4 \ ^5I_{7}$       & $4f^3 5d^1 \ ^5K_{5}$  & 0.502 & 1.892 \\
 7 & $4f^3 5d^1 \ ^5L_{7}$  &             $^5K_{6}$  & 0.681 & 2.100 \\
 8 & $4f^4 \ ^5I_{8}$       &             $^5K_{7}$  & 0.685 & 2.313 \\
 9 & $4f^3 5d^1 \ ^5K_{6}$  &             $^5I_{4}$  & 0.691 & 2.341 \\
10 &              $^5K_{7}$ &             $^5H_{3}$  & 0.915 & 2.382 \\
\hline
\end{tabular}
\caption{\bfedit{A list of the first 10 energy levels of Nd~III obtained from
the LANL {\tt CATS} code and the NIST Atomic Spectra Database.}}
\label{tab:energy_nd3}
\end{minipage}
\end{table}

\begin{table}[h]
\begin{minipage}{6.5in}
\begin{tabular}{|r|r|r|c|c|}
\hline
Level  & LANL  & NIST  & & \\
number & Label & Label & LANL (eV) & NIST (eV) \\
\hline
 1 & $4f^3 \ ^4I_{9/2}$ & $4f^3 \ ^4I_{9/2}$ &0.000 & 0.000 \\
 2 &       $^4I_{11/2}$ &       $^4I_{11/2}$ & 0.254 & 0.233 \\
 3 &       $^4I_{13/2}$ &       $^4I_{13/2}$ & 0.523 & 0.479 \\
 4 &       $^4I_{15/2}$ &       $^4I_{15/2}$ & 0.800 & 0.733 \\
 5 &       $^4F_{3/2}$  &       $^4F_{3/2}$  & 1.582 & 1.400 \\
 6 &       $^2H_{9/2}$  &       $^4F_{5/2}$  & 1.586 & 1.527 \\
 7 &       $^4F_{5/2}$  &       $^2H_{9/2}$  & 1.725 & 1.546 \\
 8 &       $^4S_{3/2}$  &       $^4F_{7/2}$  & 1.832 & 1.647 \\
 9 &       $^4F_{7/2}$  &       $^4S_{3/2}$  & 1.844 & 1.658 \\
10 &       $^4F_{9/2}$  &       $^4F_{9/2}$  & 1.993 & 1.806 \\
\hline
\end{tabular}
\caption{\bfedit{A list of the first 10 energy levels of Nd~IV obtained from
the LANL {\tt CATS} code and the NIST Atomic Spectra Database.}}
\label{tab:energy_nd4}
\end{minipage}
\end{table}

Before leaving this section, we note that we also consider another calibrated data set that includes one additional configuration of Nd~I, $4f^3\, 5d^2\, 6s$, that was omitted from the list that was originally considered by  \cite{kasen2013}.
According to the ASD, this configuration produces levels with energies that are interspersed among the energies of low-lying levels that arise from other configurations.
So it is possible that levels arising from this additional configuration could siphon population from those levels that already exist in the model.
Also, we found that this configuration can produce significant CI with the $4f^3\, 5d\, 6s^2$ and  $4f^4\, 6s\, 6p$ configurations, which are present in the original list. 
Such CI could noticeably alter the values of radiative decay rates (or oscillator strengths).
So it is desirable to investigate whether there is any sensitivity of the light curves and spectra to the inclusion of this additional configuration.

\subsection{Radiative transfer with \textsc{SuperNu}}

We use the {\tt SuperNu 4.x}\footnote{\url{https://github.com/lanl/SuperNu}} radiative transfer code
for all comparisons, thus fixing the radiative transfer code and method and only varying the input atomic data.
{\tt SuperNu}~\citep{wollaeger2013,wollaeger2014} is a multi-dimensional, multi-frequency Implicit
Monte Carlo (IMC) code with Discrete Diffusion Monte Carlo (DDMC) optimization for optically thick regions
in space and frequency (see, \eg,~\citealt{densmore2012,abdikamalov2012,cleveland2014}).
The code has an implementation of a simplified (homologous) version of the DDMC lumped Doppler shift approach
described by~\cite{wagle2023}, which in particular supports accuracy in the spectra at early time, where
the DDMC optimization is most active.
{\tt SuperNu} has both tabular and inline opacity modes, where the tabular mode interpolates
a parsed table in density and temperature and the inline mode performs a Saha-Boltzmann calculation
for excitation and ionization populations, using line list data~\citep{fontes2020}.

For all KN simulations performed with {\tt SuperNu} here, we apply the inline capability assuming LTE.
LTE is a significant caveat for the results shown at later times (after a few days)~\citep{pognan2022a,pognan2022b}.
At late time, the ionization and electron populations are not well described by Saha-Boltzmann statistics,
and in particular significant Nd II fractions may persist~\citep{hotokezaka2021}.
Hence another caveat to the present study is that lower late-time Nd I levels may obfuscate Nd I-induced
differences in KN simulations. 

We map each data set into the {\tt SuperNu} line list data format, which has two blocks per file:
level data with energies, indexes, and statistical weights, then line data with two level indexes,
and the base-10 log of the oscillator strength (see for instance, the hydrogen file {\tt data.atom.h1}
included in the {\tt SuperNu} repository).
The level data blocks are also concatenated per ion stage to the {\tt data.ion} file, which is used
by {\tt SuperNu} to construct partition functions needed for the inline Saha ionization solver.

\todo{Finish this by stating which {\tt SuperNu} opacity modes are used---are we going to use tabular?}

\section{Numerical results}
\label{sec:numres}

In the following sections, we compare the data sets obtained from each atomic physics code, with the
significant caveat that these data each represent one of potentially many possible approximation options
in each code (see, for instance,~\cite{fontes2020}, where opacities are calculated with either Dirac or
semi-relativistic Schr\"odinger equations).
However, the data used here should be reasonable representations of typical configurations and run modes
for the respective codes.

For KN light curves and spectra, we simulate a 1D spherically symmetric model with the semi-analytic
ejecta profile described by~\cite{wollaeger2018}, and used by~\cite{fontes2020,fontes2023}.
We use the power law heating rate given by Equation 18 of~\cite{wollaeger2018}, with a thermalization
efficiency of 0.25.
The composition is 100\% Nd, and the mass and mean velocity are set to 0.014 M$_{\odot}$ and 0.125$c$, 
respectively.
For all {\tt SuperNu} simulations, we use 64 uniform spatial (velocity) cells, 200 logarithmic time
steps from 10,000 s to 20 days, 1024 logarithmic wavelength bins from 1000 to 128,000~\AA, $2^{18} = 262,144$ source particles per time step, and DDMC and opacity regrouping optical depth
thresholds per cell of 10.
We only simulate bound-bound contributions to opacity, for further simplification, but these are the dominant source of opacities in KN ejecta (\eg, \citealt{kasen2013}).

In all comparisons, we refer to the \texttt{HULLAC}-generated data of~\cite{tanaka2020} as ``JLG'' (Japan-Lithuania Group),
the data from the LANL suite of relativistic atomic physics codes is referred to as ``LANL'' with
the oscillator strength lower threshold for inclusion specified parenthetically as $f_c$, and data
from \texttt{Autostructure} as ``\texttt{Autostructure}''.
The comparisons of the KN simulation data omit LANL with $f_c=10^{-6}$, as it is qualitatively consistent
with LANL $f_c=10^{-3}$.
We note we have also applied an oscillator strength threshold of $f_c=10^{-9}$ for the JLG data, incurring
a slight discrepancy between the presented tabulation and the original data.

\subsection{Some simple data metrics}
\label{sec:metrics}

Basic statistics of the atomic data may inform the subsequent comparisons of light curves and spectra.
In Table~\ref{tab2:nlines} are the number of lines and levels per ion stage (rows) and per data set
(columns).
We see that there is significant variation in the number of levels and lines in the neutral stage, with the
LANL data having the highest number of levels and the highest number of lines for an oscillator strength
cutoff of $f_c=10^{-6}$.
The number of levels in the neutral ion stage are comparable between LANL and \texttt{Autostructure} data, whereas
the JLG data has $\sim6-9\times$ fewer levels.
However, the number of levels for the higher ion stages are comparable among all data sets.
For LANL data with $f_c=10^{-6}$, JLG and \texttt{Autostructure}, the number of lines for Nd II-IV are comparable
in order of magnitude.
\todo{Fontes: anything worth mentioning on how differences in the \# of level/lines manifest among codes?}

\begin{table}[!h]
    \centering
    \begin{tabular}{|c|c|c|c|c|}
        \hline
        (\# levels / \# lines) & LANL ($f_c=10^{-6}$) & LANL ($f_c=10^{-3}$) & JLG & \texttt{Autostructure} \\
        \hline
        Nd I & {\bf (18,104 / 19,116,842)} & {\bf (18,099 / 815,431)} & (2,189 / 353,548) & {\bf (12,215 / 4,682,375)} \\
        Nd II & (6,888 / 3,390,966) & (6,888 / 375,026) & {\bf (5,249 / 2,242,145)} & (6,888 / 3,842,130) \\
        Nd III & (1,650 / 197,010) & (1,649 / 33,313) & (1,630 / 224,049) & (2,252 / 455,542) \\
        Nd IV & (241 / 5,276) & (241 / 2,155) & (390 / 15,453) & (474 / 23,849) \\
        \hline
    \end{tabular}
    \caption{Number of levels and lines per data set, per ion stage.
        Maximum values per column are in bold.
        LANL data is shown for two oscillator strength lower bounds.
        The number of levels in Nd I is comparable between LANL and \texttt{Autostructure} data sets, but significantly
        lower in the JLG data set.
        The number of lines in Nd I is significantly different across all data sets.}
    \label{tab2:nlines}
\end{table}

While the number of levels and lines indicate differences, it cannot be discerned from those numbers alone
how KN light curves and spectra are affected.
For instance, a data set with a large number of lines can have many weak lines with low oscillator strengths.
Consequently, we find it useful to examine a few other simple metrics that may better elucidate the effects
of the data set on KN emission.
For instance, Table~\ref{tab3:metrics} shows the geometric average and standard deviation of $gf$, the lower
statistical weight multiplied by the oscillator strength of the lines in an ion stage.
We compute the average as
\begin{equation}
    \log_{10}(gf) = \frac{1}{N_l}\sum_{l=1}^{N_l}\log_{10}(g_lf_l) \;\;,
\end{equation}
where $N_l$ is the number of lines,
$l$ is a line index, and $g_l$ and $f_l$ are the lower statistical weight and oscillator strength of line $l$,
respectively. This quantity $gf$~is often called the symmetric weighted oscillator strength: For an absorption process from energy level 1 to 2 and the inverse emission from 2 to 1, $|g_{1} f_{12}| = |g_2 f_{21}|$. 
The formula for the standard deviation of $gf$~is 
\begin{equation}
    \hat{\sigma}_{gf}
    = \frac{1}{|\log_{10}(gf)|}
    \sqrt{\frac{1}{N_l-1}\sum_{l=1}^{N_l}\left(\log_{10}\left(\frac{g_lf_l}{gf}\right)\right)^2}
    \;\;.
\end{equation}
The standard deviation is normalized by the corresponding average to facilitate comparison between data sets. Thus, a standard deviation of 0.3 is 30\% of the average.
As expected, the LANL data set with $f_c=10^{-3}$ have higher averages than the LANL data set with $f_c=10^{-6}$
in each ion stage.
The averages for Nd I-II are comparable between LANL ($f_c=10^{-6}$), JLG and \texttt{Autostructure}, but JLG has relatively
high averages for Nd III-IV.
Given the comparable number of levels and lines, this discrepancy indicates that there are differences
in the distribution of number of lines per interval of $g_lf_l$ values.
The fractional standard deviations are similar between \texttt{Autostructure} and the two LANL data sets; they show a gradual
increase with increasing ionization.
In contrast, the JLG data set shows a sharper increase in fractional standard deviation with increasing ionization.
While this metric shows atomic physics approximation differences, the effect in KN spectra does not directly
follow, since the metric does not factor in excitation populations.

\begin{table}[!h]
    \centering
    \begin{tabular}{|c|c|c|c|c|}
        \hline
        ($\log_{10}(gf)$ / $\hat{\sigma}_{gf}$)
        & LANL ($f_c=10^{-6}$) & LANL ($f_c=10^{-3}$) & JLG & \texttt{Autostructure} \\
        \hline
        Nd I & (-3.6 / 0.2) & (-1.7 / 0.3) & (-4.2 / 0.4) & (-4.0 / 0.4) \\
        Nd II & (-3.3 / 0.3) & (-1.6 / 0.3) & (-2.9 / 0.6) & (-3.8 / 0.4) \\
        Nd III & (-3.2 / 0.3) & (-1.6 / 0.3) & (-1.5 / 1.1) & (-3.7 / 0.4) \\
        Nd IV & ({\bf -2.5} / 0.5) & ({\bf -1.2} / 0.6) & ({\bf -0.6} / 2.6) & ({\bf -2.9} / 0.5) \\
        \hline
    \end{tabular}
    \caption{Geometric average and fractional standard deviation of lower level statistical weight times
        line oscillator strength per data set, per ion stage (two significant digits).
        The standard deviation values are divided by the average.
        Maximum average values per column are in bold.
        LANL data is shown for two oscillator strength lower bounds.
        JLG has both relatively high averages for Nd III-IV and a sharp increase in fractional standard
        deviation with ion stage.
        With the comparable number of lines in these ion stages, this indicates a difference in the number of
        lines per interval over $g_l f_l$.}
    \label{tab3:metrics}
\end{table}

A basic measure of average line wavelength per ion stage, per data set, may be useful in setting an
expectation of where the bulk of the spectrum might be at particular times in the KN evolution.
This metric should factor in line strength, as vanishingly small lines should produce little absorption or emission.
Table~\ref{tab4:weightedwavelens} has oscillator strength-weighted average line wavelength per data set, per ion stage\bfedit{, in microns ($\mu$m)}.
The wavelength averages decrease with increasing ionization for all atomic data sets.
We see that both LANL data sets show a relatively longer average wavelength in the neutral stage.
This discrepancy in Nd I corresponds to mid-infrared lines with high oscillator strengths in
the LANL data set, unique to Nd in terms of their strength~\citep{fontes2020,even2020}.
These mid-IR Nd I lines are resonances between the partially filled $4f^4$ ground
configuration and corresponding excited states~\citep{even2020}, and have been shown to be sensitive
in structure and location to the atomic physics model (see, for instance, the comparison of fully- and semi-relativistic atomic physics models of~\citealt{fontes2020}). \todo{Explain why here... get Fontes thoughts on this.}

\begin{table}[!h]
    \centering
    \begin{tabular}{|c|c|c|c|c|}
        \hline
        $\frac{1}{\sum_{l=1}^{N_l}f_l}\sum_{l=1}^{N_l}f_l\lambda_l$ & LANL ($f_c=10^{-6}$) & LANL ($f_c=10^{-3}$) &
        JLG & \texttt{Autostructure} \\
        \hline
        Nd I & {\bf 1.00} & {\bf 0.76} & {\bf 0.46} & {\bf 0.49} \\
        Nd II & 0.46 & 0.42 & 0.35 & 0.48 \\
        Nd III & 0.26 & 0.26 & 0.15 & 0.23 \\
        Nd IV & 0.20 & 0.19 & 0.09 & 0.20 \\
        \hline
    \end{tabular}
    \caption{Oscillator strength-weighted average of line wavelength (\textmu m) per data set, per ion stage. Maximum values per column are in bold. LANL data is shown for two oscillator strength lower bounds. Agreement is good among the four data sets for Nd II-IV, but the LANL data sets have a significantly
    longer average wavelength for Nd I.}\label{tab4:weightedwavelens}
\end{table}

\subsection{Temperature and ionization in space and time}
\label{sec:teion}

\begin{figure}[!h]
    \centering
    \includegraphics[width=0.49\linewidth]{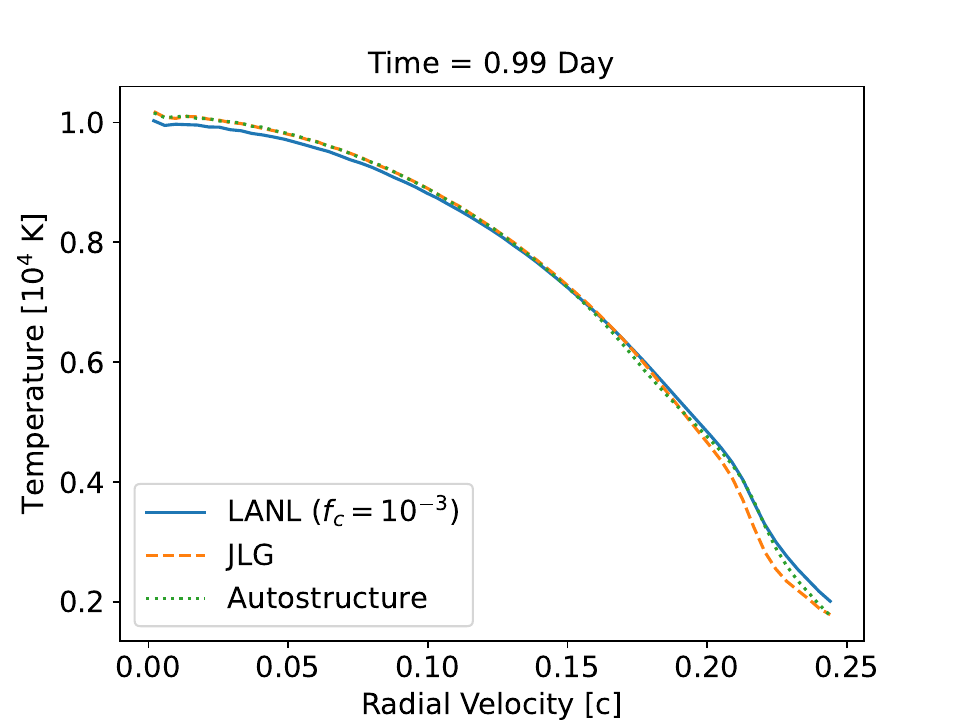}
    \includegraphics[width=0.49\linewidth]{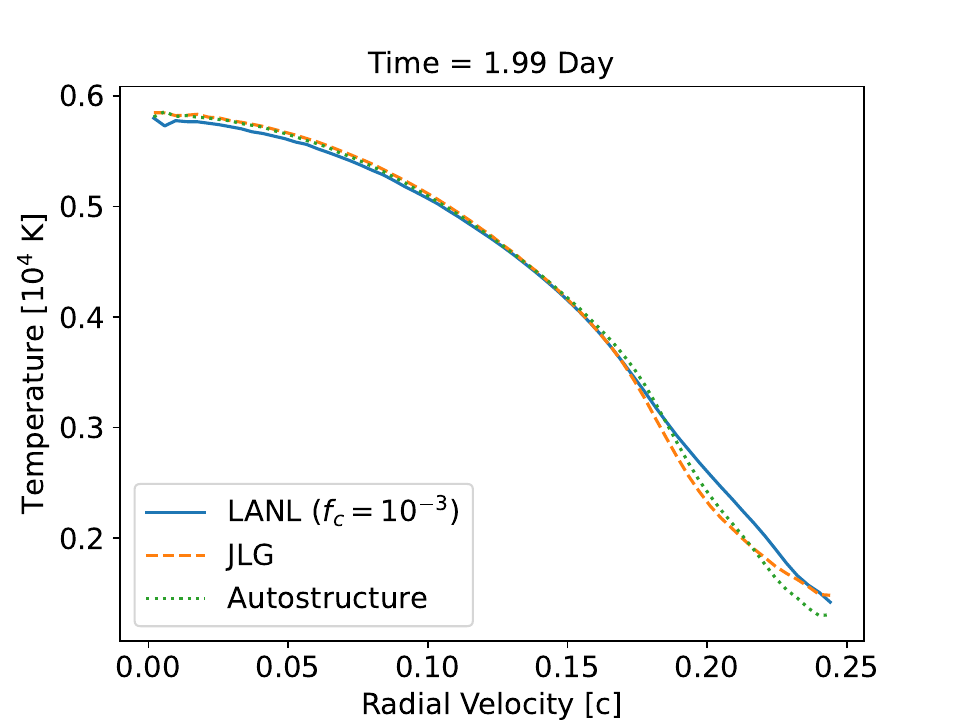} \\
    \includegraphics[width=0.49\linewidth]{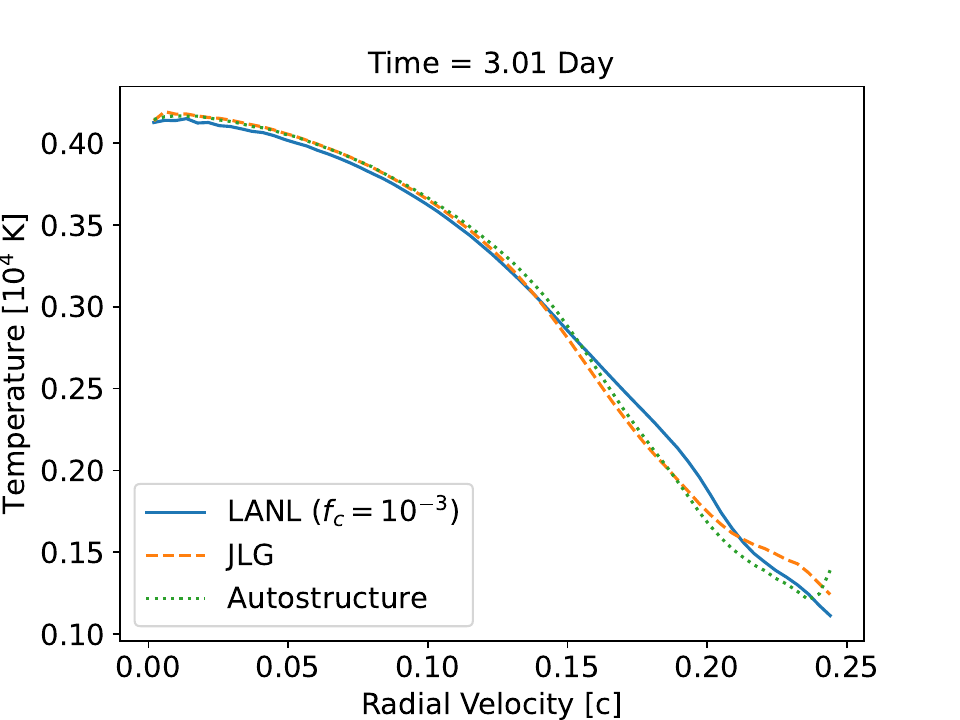}
    \includegraphics[width=0.49\linewidth]{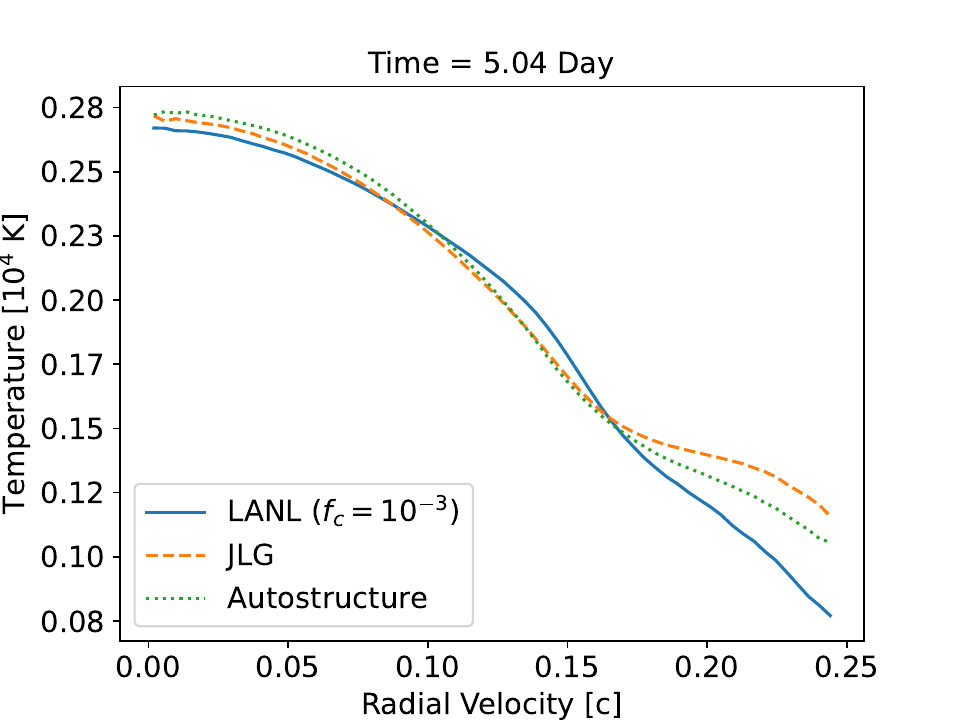} \\
    \includegraphics[width=0.49\linewidth]{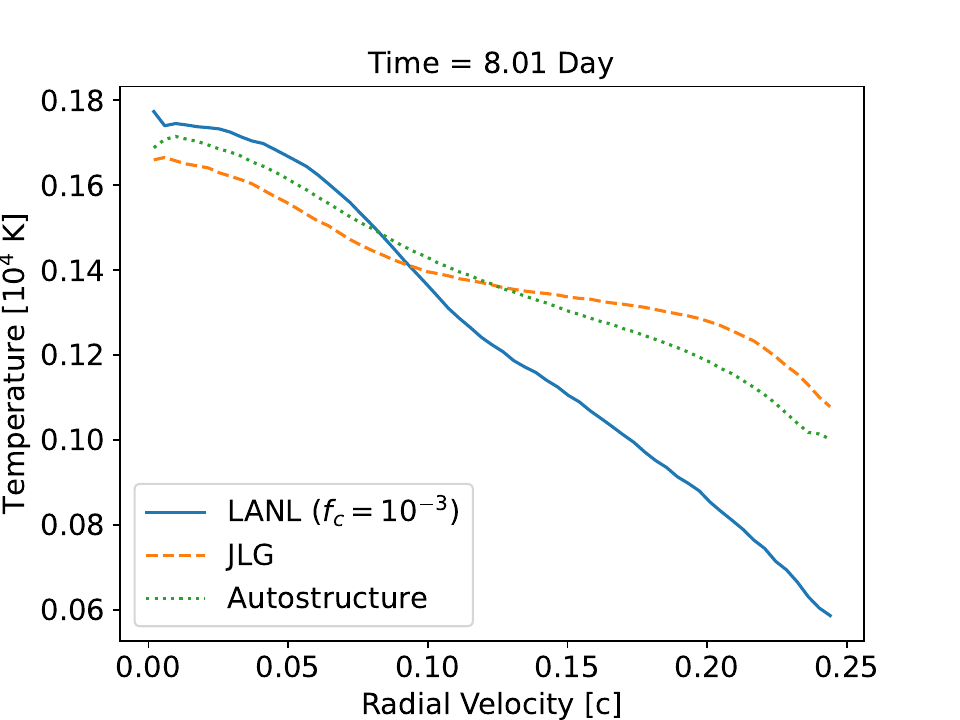}
    \includegraphics[width=0.49\linewidth]{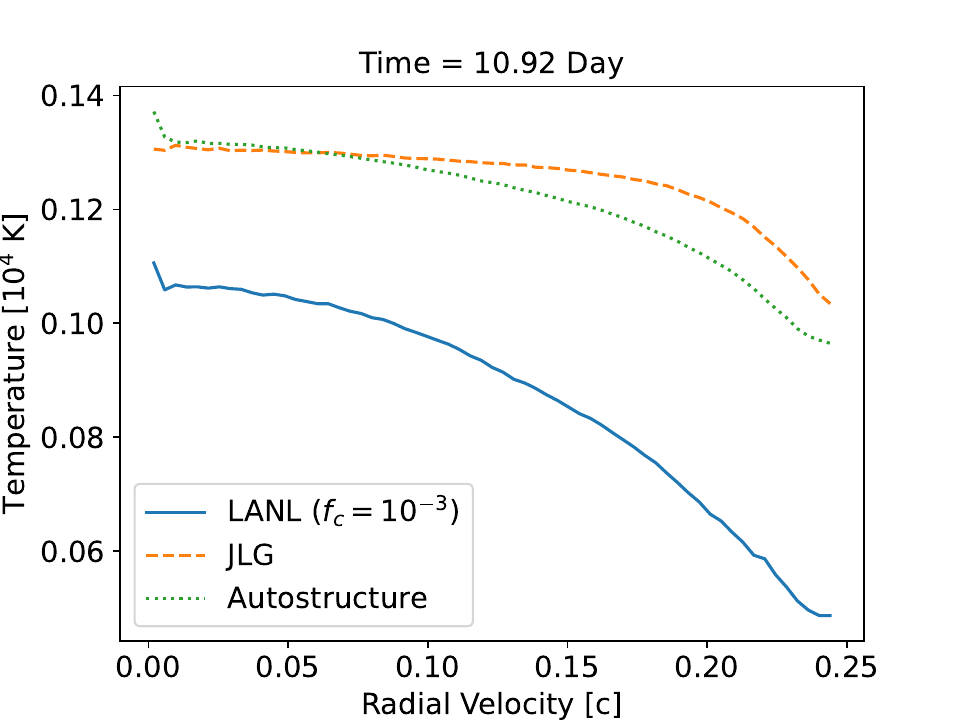}
    \caption{Matter temperature versus velocity coordinate at day 1, 2, 3, 5, 8, and 11 (left to right,
    top to bottom) for the KN simulation, using LANL (solid), JLG (dashed) and \texttt{Autostructure} (dotted).
    The temperature profiles produced by all three data sets are in close agreement until day 5, when
    they start to deviate at outer radii.
    This discrepancy between calculations moves inward with time, following the inward-moving wave of Nd II recombining to Nd I.}
    \label{fig1:teion}
\end{figure}

We now examine the impact of the data sets on the state of matter in the {\tt SuperNu} simulations.
Given the differences in data discussed in the previous section, it is reasonable to expect differences
in KN ejecta matter temperature and ionization levels in both space and time.
Matter temperature versus velocity coordinate is shown for days 1, 2, 3, 5, 8 and 11 (left to right, top
to bottom) in Figure~\ref{fig1:teion}.
The temperatures are initially in good agreement among the three data-set results, up to approximately
day 5.
After day 5, the \texttt{Autostructure} (dotted) and JLG (dashed) results have high temperatures at outer radii,
and the temperature difference grows toward inner radii from day 5 to day 11.

\begin{figure}[!h]
    \centering
    \includegraphics[width=0.49\linewidth]{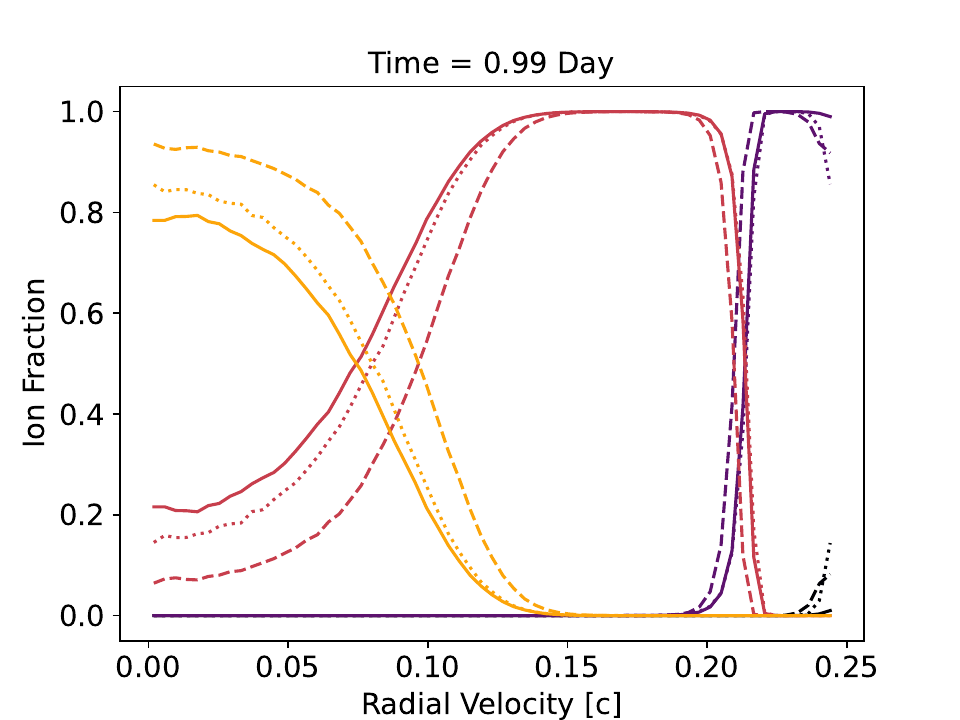}
    \includegraphics[width=0.49\linewidth]{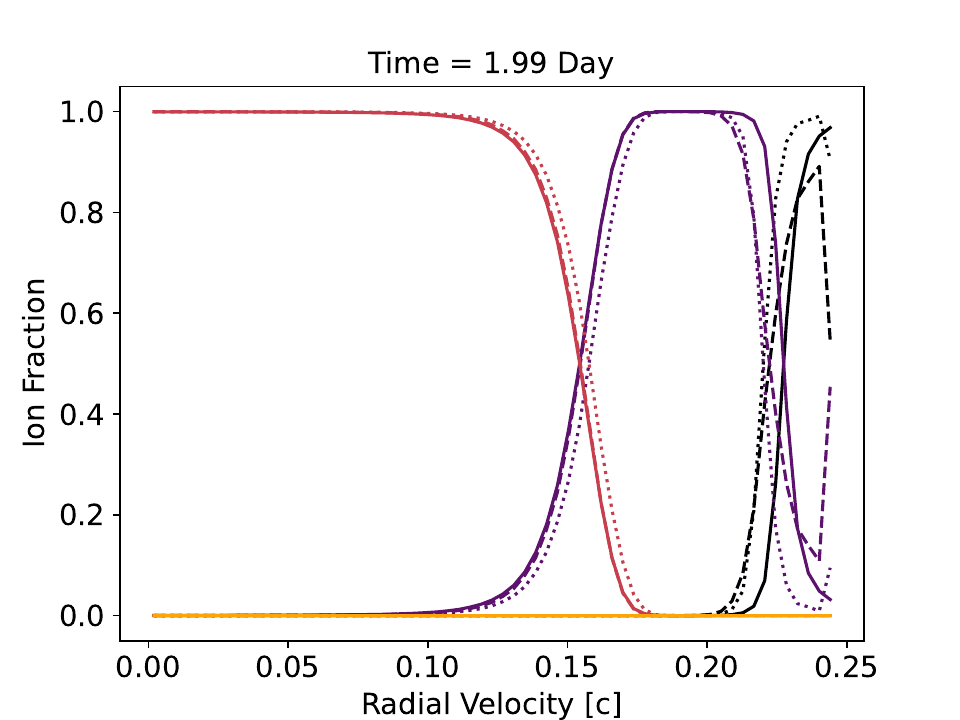} \\
    \includegraphics[width=0.49\linewidth]{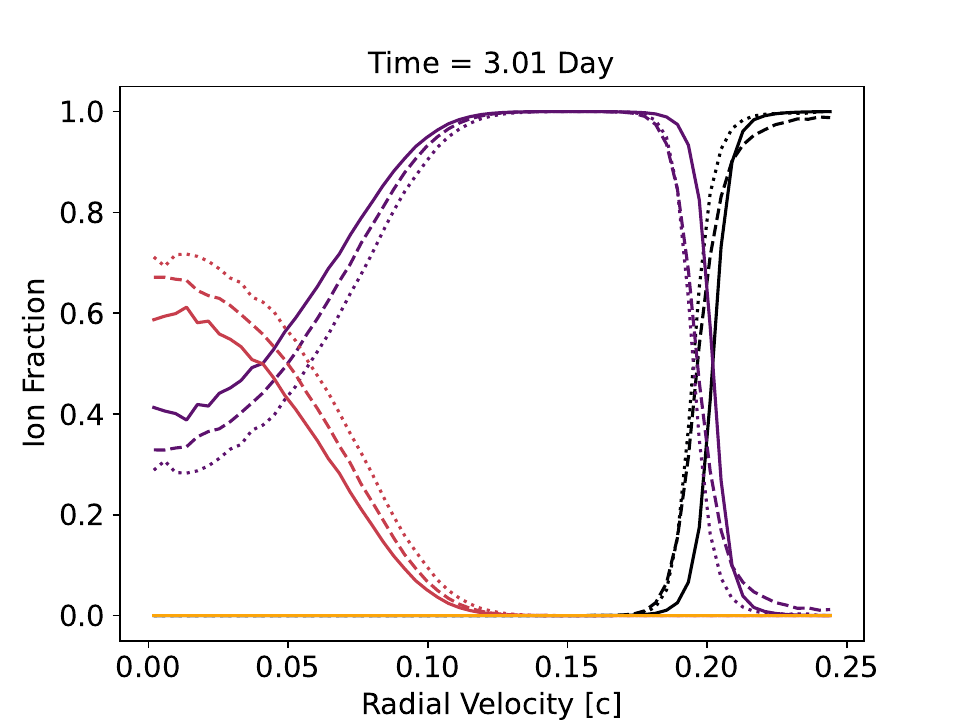}
    \includegraphics[width=0.49\linewidth]{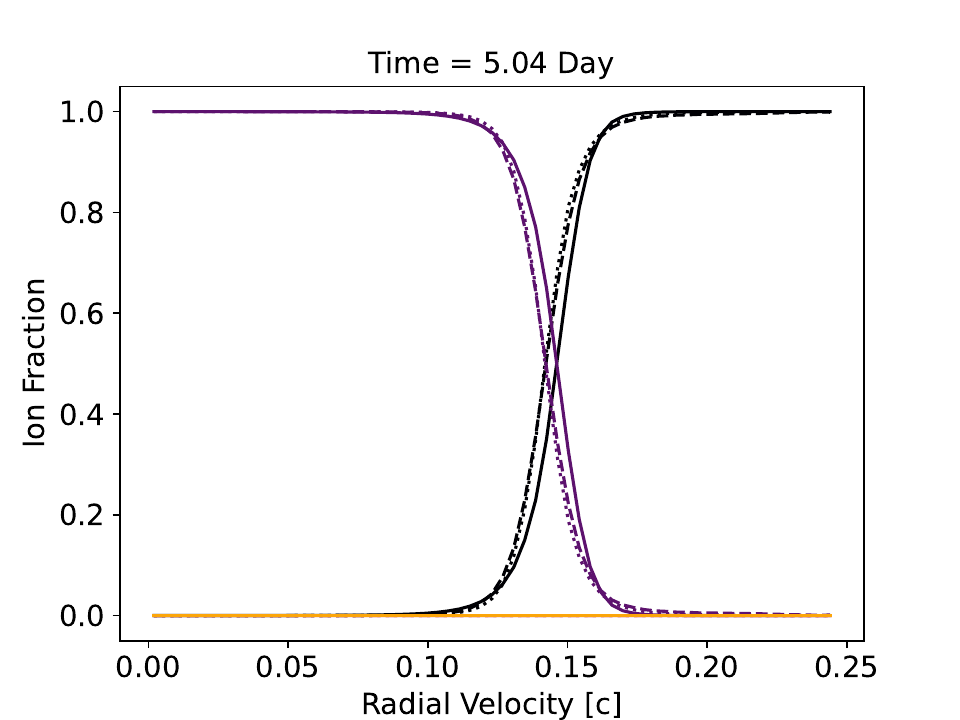} \\
    \includegraphics[width=0.49\linewidth]{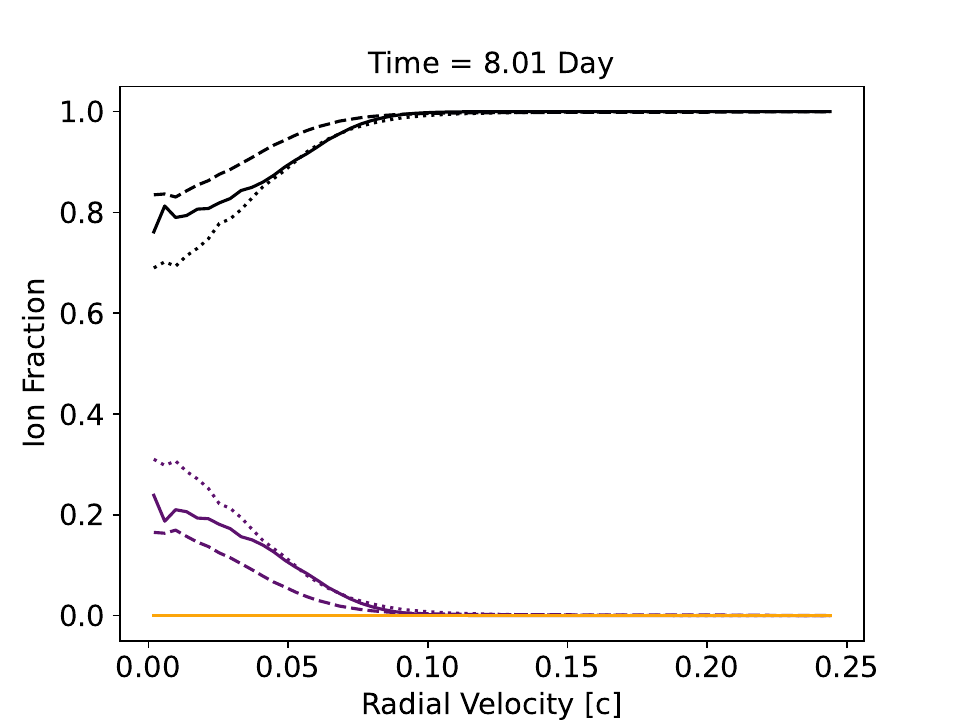}
    \includegraphics[width=0.49\linewidth]{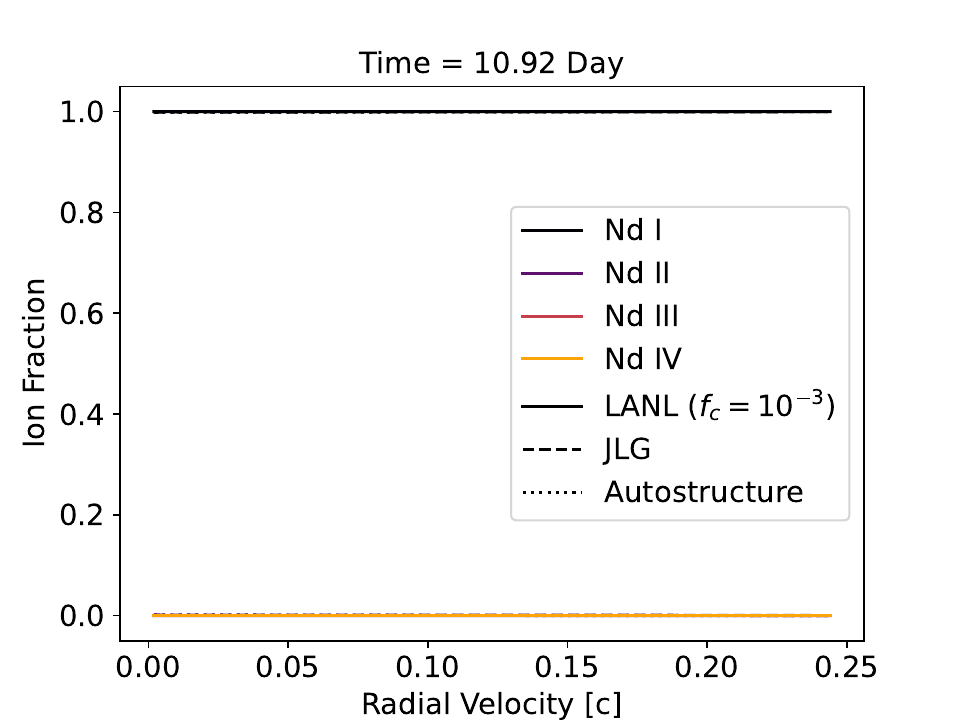}
    \caption{Ionization fraction versus velocity coordinate at day 1, 2, 3, 5, 8, and 11 (left to right,
    top to bottom) for the KN simulation, using LANL (solid), JLG (dashed) and \texttt{Autostructure} (dotted).
    Ionization trends agree throughout time for all three data sets: all four ion stages are present by day 1
    and by day 5 only Nd II and Nd I remain, with Nd II recombining to Nd I.
    All results are computed with the Saha (LTE) equation: non-LTE effects may strongly impact these trends,
    especially Nd I dominance at late time (see, for instance,~\citealt{hotokezaka2021,pognan2022a,brethauer2025}).}
    \label{fig2:teion}
\end{figure}

Figure~\ref{fig2:teion} has the population fraction versus velocity coordinate for the four ion stages
for days 1, 2, 3, 5, 8 and 11 (left to right, top to bottom) from the Nd KN simulation, using LANL with
$f_c=10^{-3}$ (solid), JLG (dashed) and \texttt{Autostructure} (dotted) data sets.
The ionization trends are similar among the three data sets used, with Nd III and IV fractions agreeing
closely between LANL and \texttt{Autostructure} data sets at inner radii at day 1.
For all data sets, at day 2 Nd III dominates at inner radii ($v \lesssim 0.15c$) while Nd II begins to
recombine to Nd I at outer radii.
From day 3 to day 5, Nd III vanishes at the inner radii, recombining to Nd II, and Nd II progressively
recombines to Nd I at decreasing outer radii.
By about day 8, Nd I dominates the KN ejecta for all data sets used.
Comparing the trend in ionization data to the trend in temperature curves in Figure~\ref{fig1:teion}, we
note that the temperature for the LANL data set starts to significantly diverge (around day 5) over a
range of outer radii comparable to the radii over which Nd I dominates.

\subsection{Light curves and spectra}
\label{sec:lcspec}

\begin{figure}[!h]
    \centering
    \includegraphics[width=0.49\linewidth]{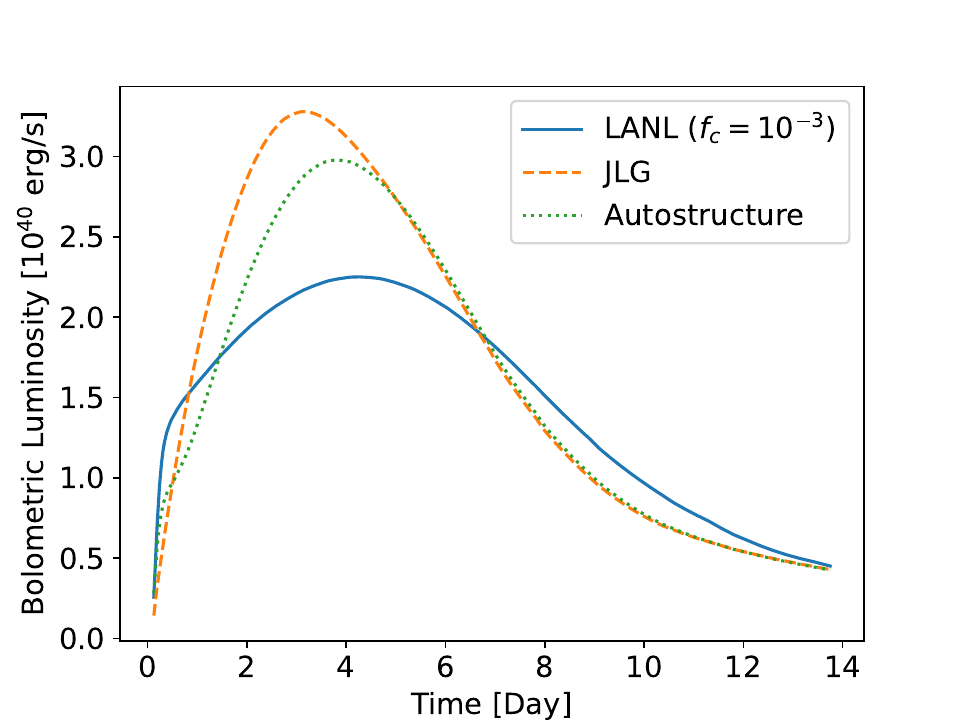}
    \includegraphics[width=0.49\linewidth]{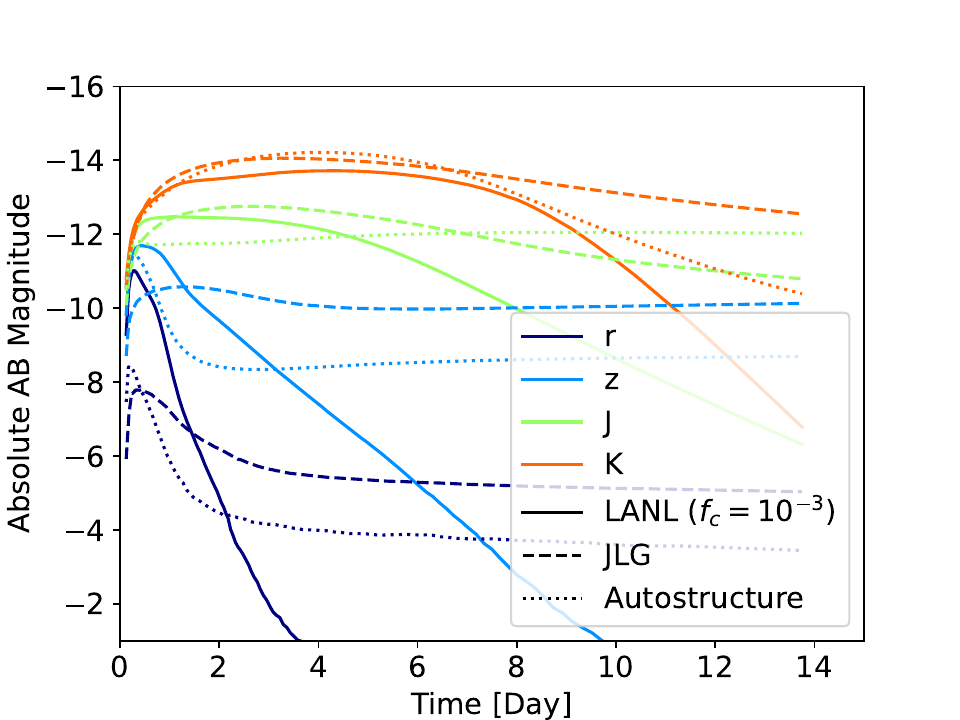}
    \caption{Bolometric luminosity (left) and broadband magnitudes (right) versus time, for LANL (solid),
    JLG (dashed) and \texttt{Autostructure} (dotted) data sets.
    Bolometric luminosities are comparable in order of magnitude; optical band magnitudes notably decay
    at distinct rates depending on the data set, with the LANL data set showing the steepest decline.}
    \label{fig1:lcspec}
\end{figure}

Bolometric luminosity versus time (left) and absolute AB magnitudes for $r$, $z$, $J$, and $K$~filters 
versus time (right) are shown in Figure~\ref{fig1:lcspec} for the three data sets.
The peak bolometric luminosities are consistent within a factor of $\sim$$2$~across all three atomic codes, with JLG (dashed lines)
and \texttt{Autostructure} (dotted lines) agreeing at $\sim$10\%.
The luminosity with the LANL data set (solid lines) is the outlier, both in peak luminosity and in
broadness of the light curve, indicating a higher effective opacity.
The absolute AB magnitudes disagree quantitatively between all three results.
At $\lesssim 2$ days the LANL data set produces systematically bluer emission and the tails of the light curves drop at a steeper slope.
The \texttt{Autostructure} and JLG data sets produce structurally similar light curves, but these differ by $\sim2$~AB mag in $r$~and $z$~bands at $\gtrsim 2~\mathrm{days}$.
The tails of the infrared $J$ and $K$~bands all evidently diverge; however, up to 8 days, the $K$~band light curves appear to be in good agreement.

\begin{figure}[!h]
    \centering
    \includegraphics[width=0.49\linewidth]{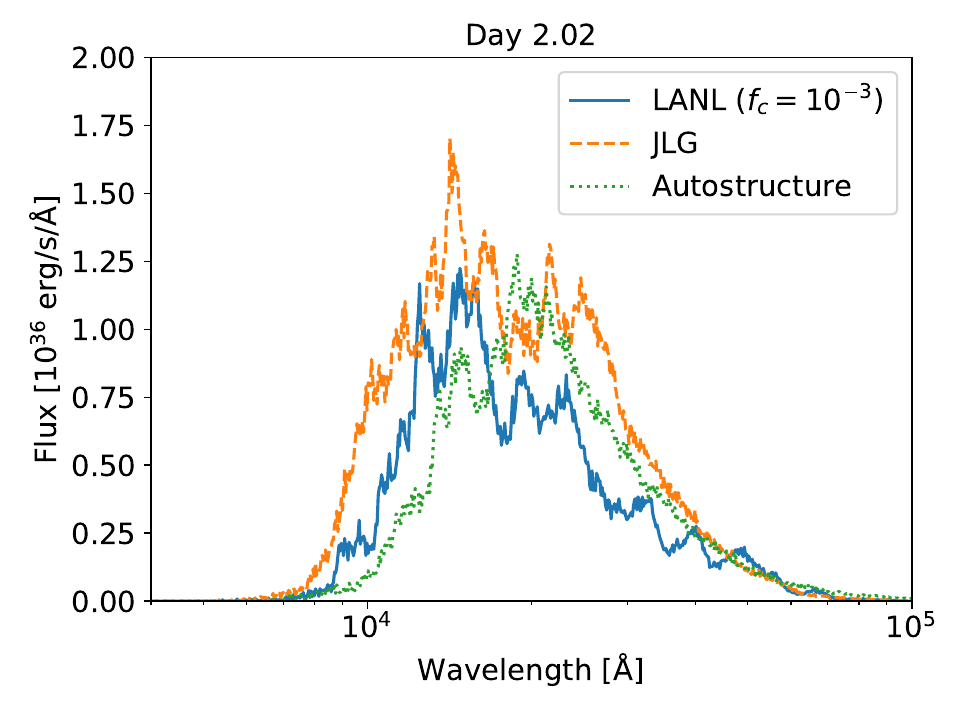}
    \includegraphics[width=0.49\linewidth]{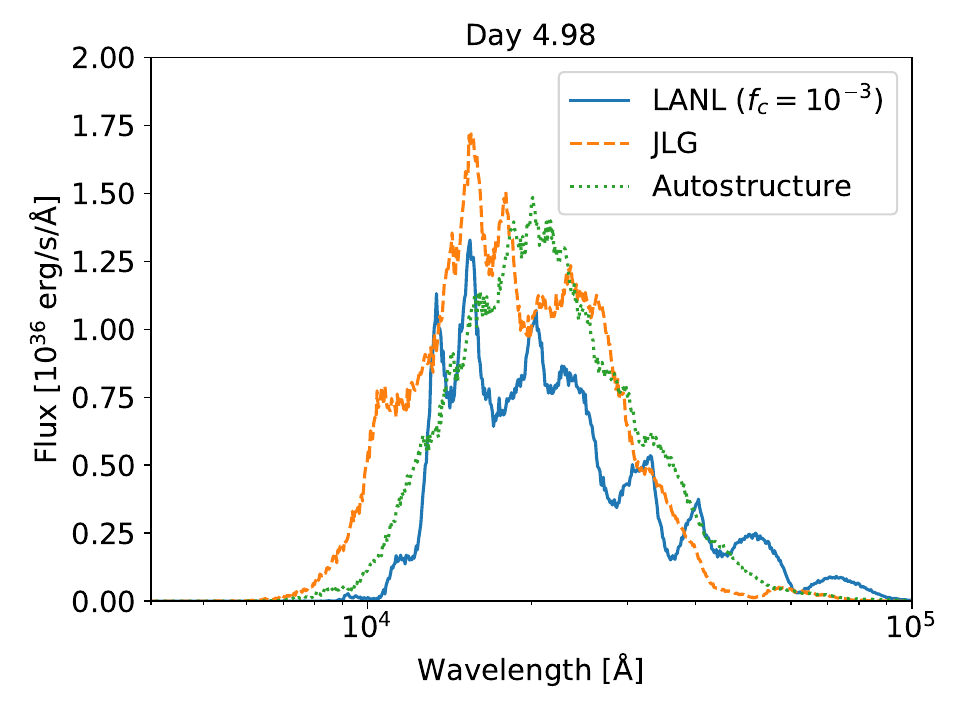} \\
    \includegraphics[width=0.49\linewidth]{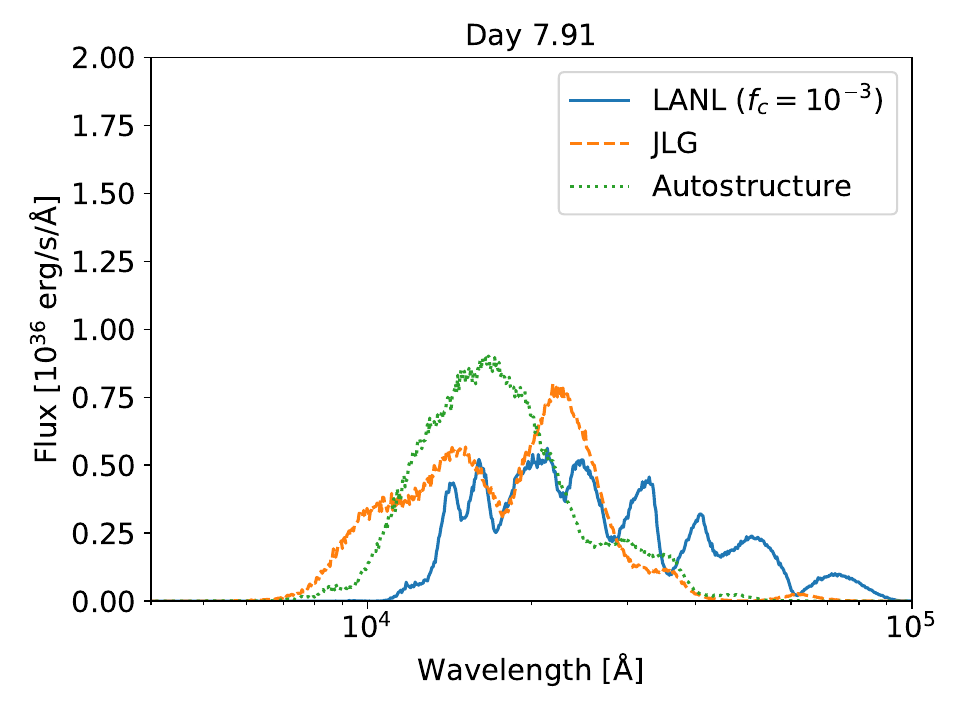}
    \includegraphics[width=0.49\linewidth]{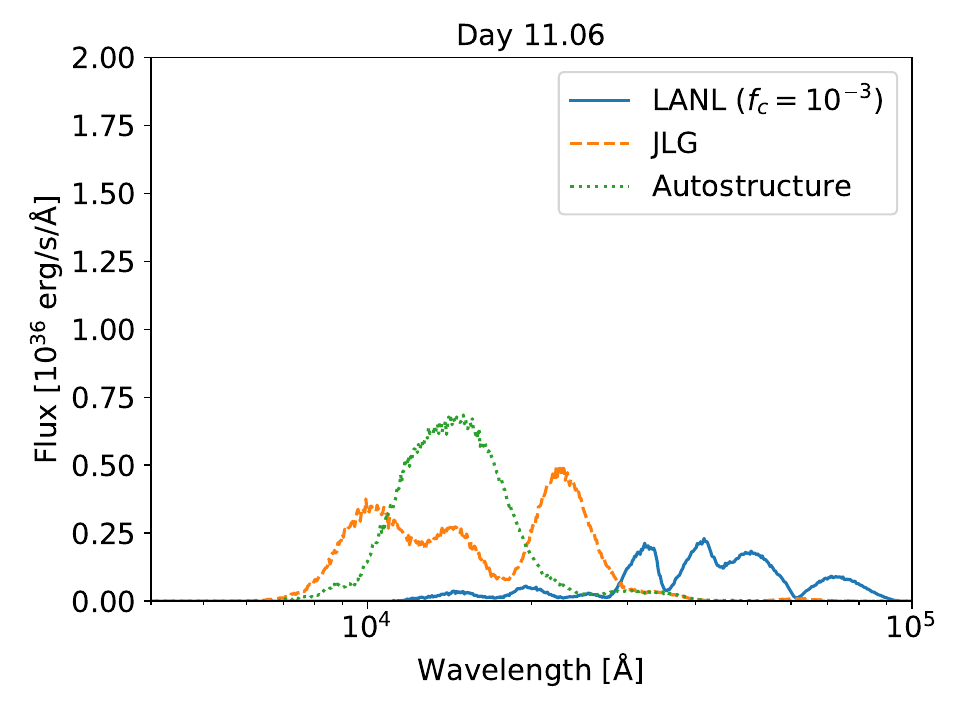}
    \caption{Spectra versus wavelength at day 2 (upper left), day 5 (upper right), day 8 (lower left)
    and day 11 (lower right), for LANL (solid), JLG (dashed) and \texttt{Autostructure} (dotted) data sets.
    Bulk emission agrees well among all three data sets at early time, but all spectra diverge at late time,
    when Nd I begins to dominate the ionization population.
    All spectra show significant discrepancy.
    The LANL data set has more line structure in the mid-IR range, which starts appearing by day 2.}
    \label{fig2:lcspec}
\end{figure}

Figure~\ref{fig2:lcspec} shows spectra versus wavelength for day 2 (upper left), day 5 (upper right), day 8 (lower left) and day 11 (lower right) for the same simulations as in Figure~\ref{fig1:lcspec}.
The spectra at different times show trends that are consistent with the broadband magnitudes, where
the LANL data set (solid) is initially comparable or brighter at low wavelengths, but over time shifts
to become significantly redder relative to either the JLG (dashed) or \texttt{Autostructure} (dotted) spectra.
At late time, Nd contributes significant mid-IR lines to opacity and emissivity~\citep{korobkin2021}. 
This shift is complemented by the growth of significant mid-IR line structure in the simulation with the LANL data set, which is apparent at day 2 but does not yet dominate the spectrum.
At day 11, there are significant differences in spectral structure among the three simulations.

\begin{figure}[!h]
    \centering
    \includegraphics[width=0.49\linewidth]{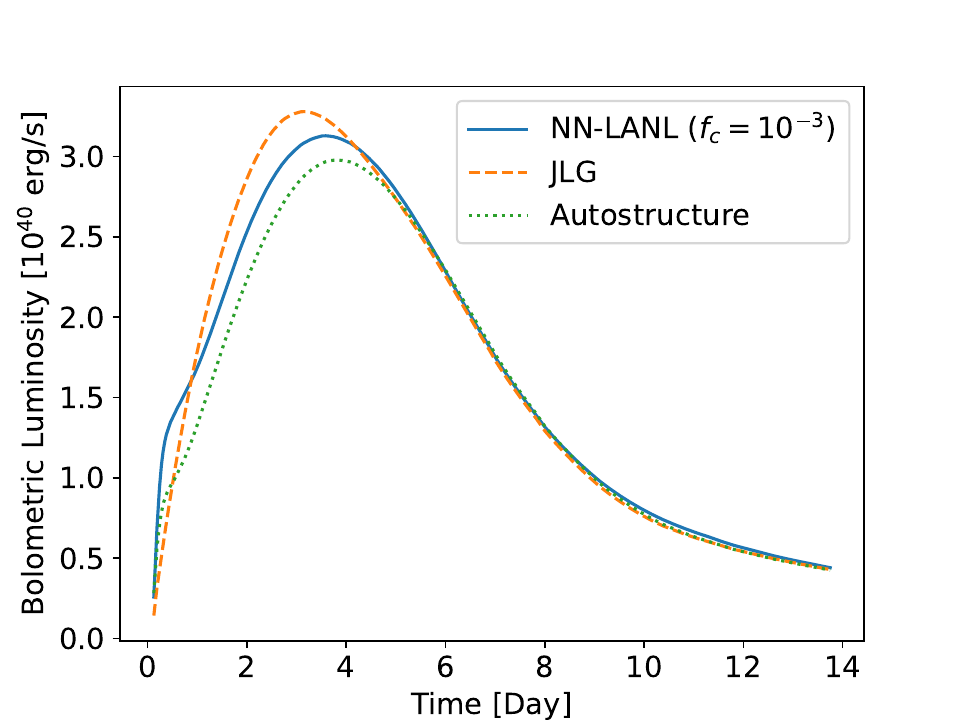}
    \includegraphics[width=0.49\linewidth]{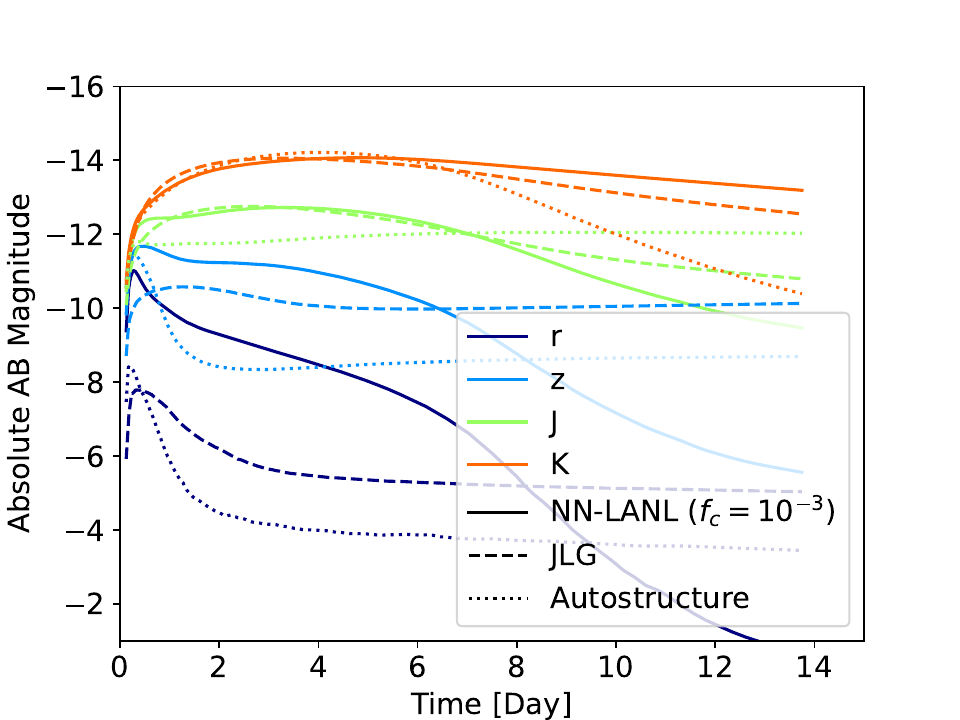}
    \caption{Same as Figure~\ref{fig1:lcspec}, but with LANL data with minimized level and line list data
    for Nd I (NN-LANL).
    The minimization of Nd I causes the LANL bolometric luminosity to agree much more closely with that of the
    JLG and \texttt{Autostructure} data sets.
    The rate of decline in the optical band magnitudes is reduced, from the full LANL data set, but still
    not as shallow as \texttt{Autostructure} and JLG data sets.}
    \label{fig3:lcspec}
\end{figure}

The significant spectral shift of the simulation with the LANL data set at late time, in concert with
the dominance of Nd I at late time observed in Figure~\ref{fig2:teion} and the strong lines observed in the
late-time Nd opacity~\citep{korobkin2021}, motivates a simulation with LANL data, but with a reduced level
and line list for the neutral stage of Nd.
In particular, we replace the Nd I line list file of LANL ($f_c=10^{-3}$) ({\tt data.atom.nd1}) with
an artificial file containing two levels and one line with vanishingly small oscillator strength (and statistical
weight 1).
We ensure the partition function used in the Saha ionization has the same two levels for Nd I.
Results from simulations with this modified data are labeled ``NN-LANL'' (NN = ``no neutral'').
In Figure~\ref{fig3:lcspec} we plot the same light curves as in Figure~\ref{fig1:lcspec}, but with LANL data replaced by NN-LANL results.
The bolometric luminosity for NN-LANL data is in closer agreement with the JLG and \texttt{Autostructure} data,
relative to that of the unmodified LANL data.
We note also that the tails of the $z$ and $r$-band curves are broader in time, compared to those of
the original LANL data set.
However, the broader tails still eventually trend below the relatively flat $r$ and $z$-band tails
of the JLG and \texttt{Autostructure} data sets, at $\gtrsim 8~\mathrm{days}$.

\begin{figure}[!h]
    \centering
    \includegraphics[width=0.49\linewidth]{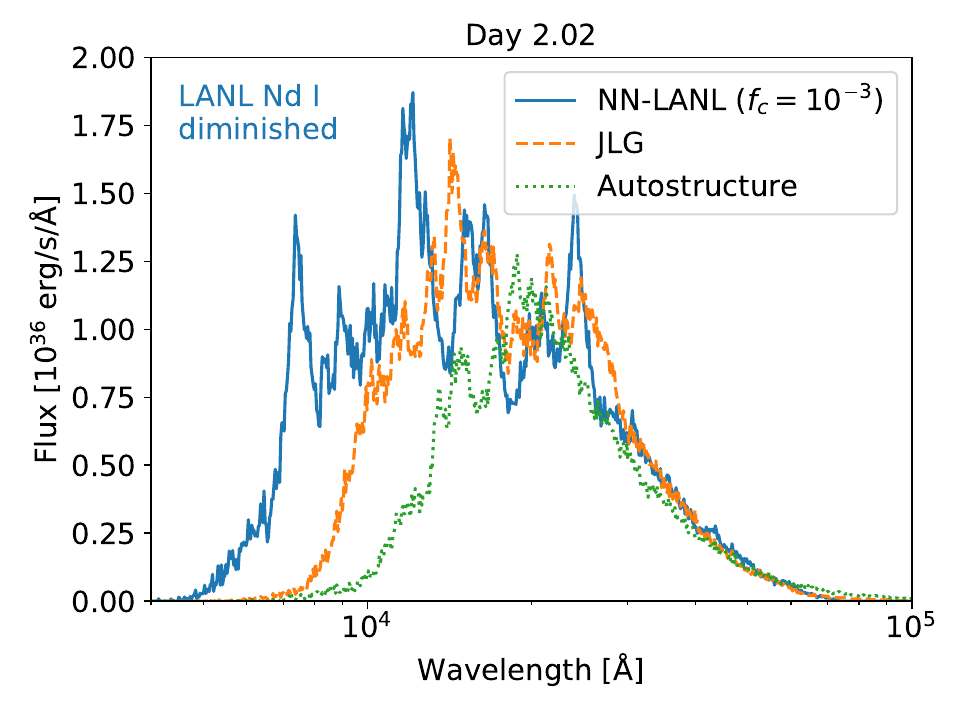}
    \includegraphics[width=0.49\linewidth]{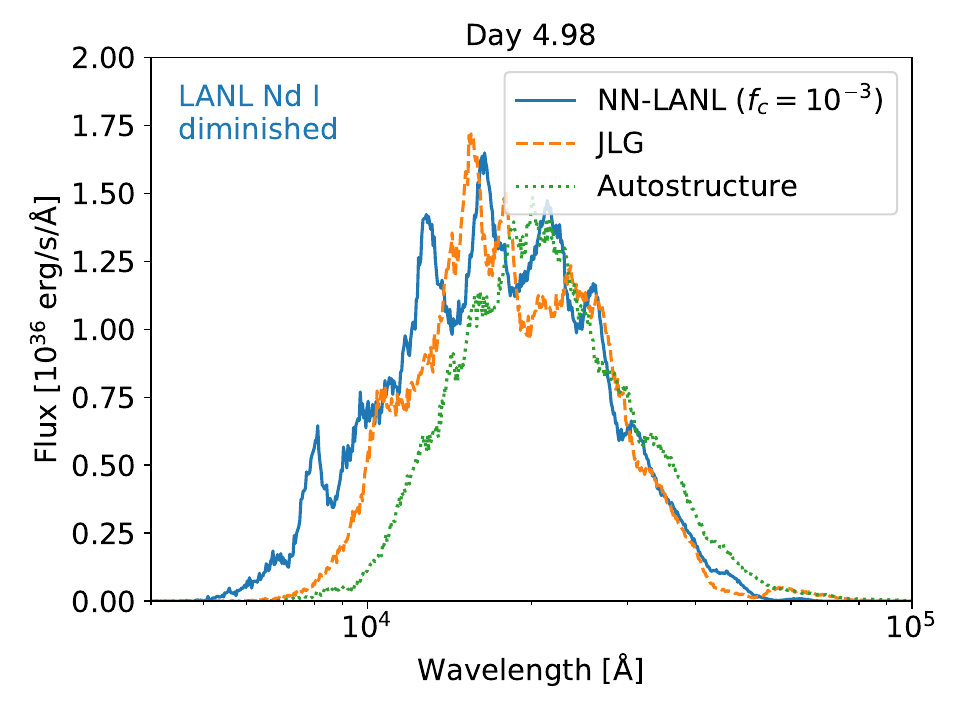} \\
    \includegraphics[width=0.49\linewidth]{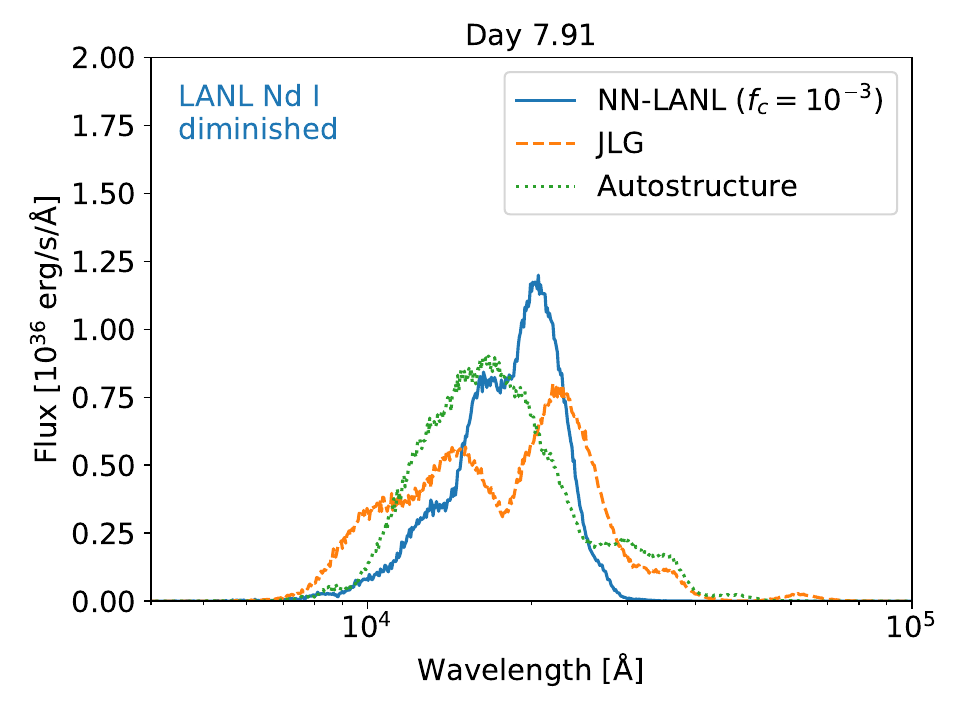}
    \includegraphics[width=0.49\linewidth]{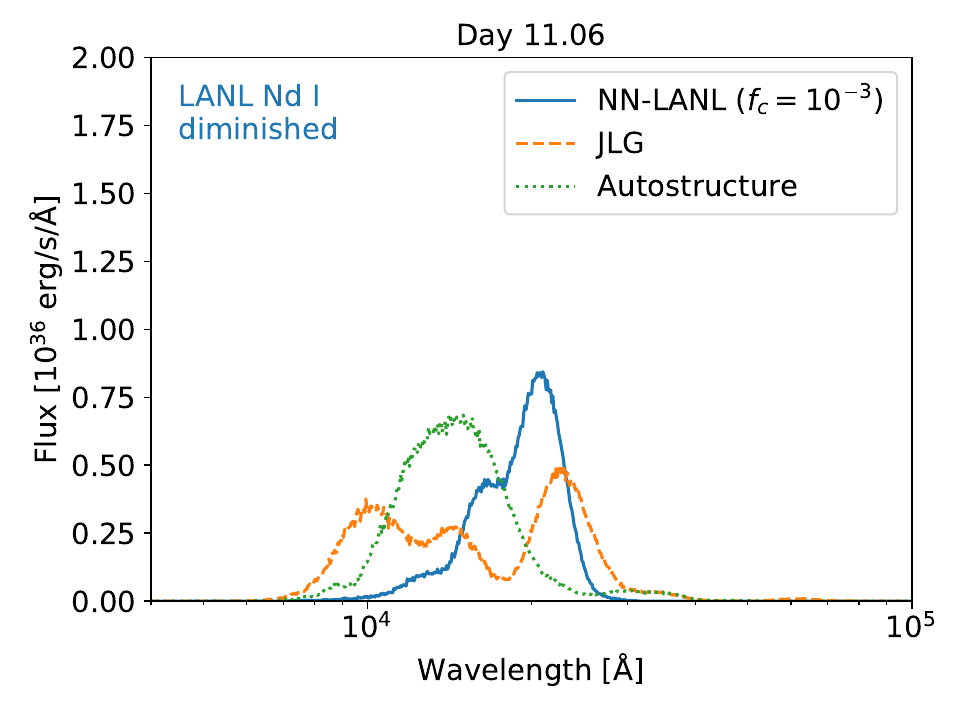}
    \caption{Same as Figure~\ref{fig2:lcspec}, but with LANL data with minimized level and line list data
    for Nd I (NN-LANL).
    At day 2, we see that NN-LANL data produces a bluer spectrum between 1000 and 10000~\AA.
    Additionally, much of the mid-IR line structure is removed, resembling the spectra of the other data sets.}
    \label{fig4:lcspec}
\end{figure}

The spectra in Figure~\ref{fig4:lcspec} are produced with the same data as Figure~\ref{fig2:lcspec}, but use the NN-LANL calculations in place of the original LANL data.
We see better agreement of the NN-LANL data with the other two data sets in the location of the bulk emission.
Removing the lines from Nd I in the LANL data set removes the strong lines in the mid-IR range of wavelength,
making the NN-LANL result relatively featureless as of $\gtrsim 8~\mathrm{days}$.
We also note that the NN-LANL data set furnishes better agreement with the JLG and \texttt{Autostructure} data sets in
late-time matter temperature, consistent with the bluer bulk emission.

\begin{figure}[!h]
    \centering
    \includegraphics[width=0.49\linewidth]{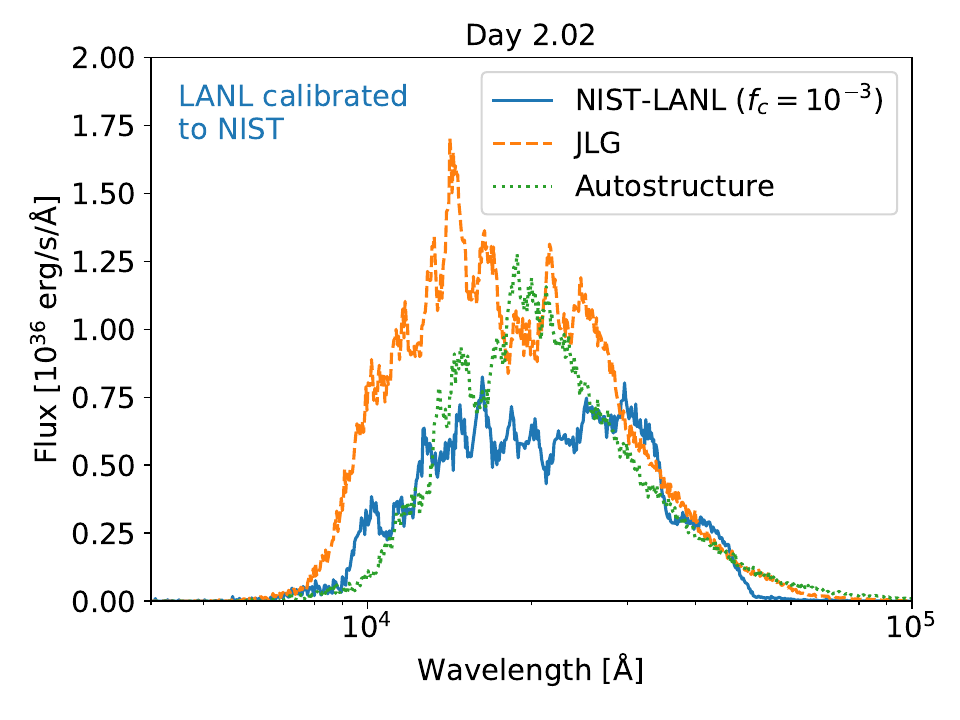}
    \includegraphics[width=0.49\linewidth]{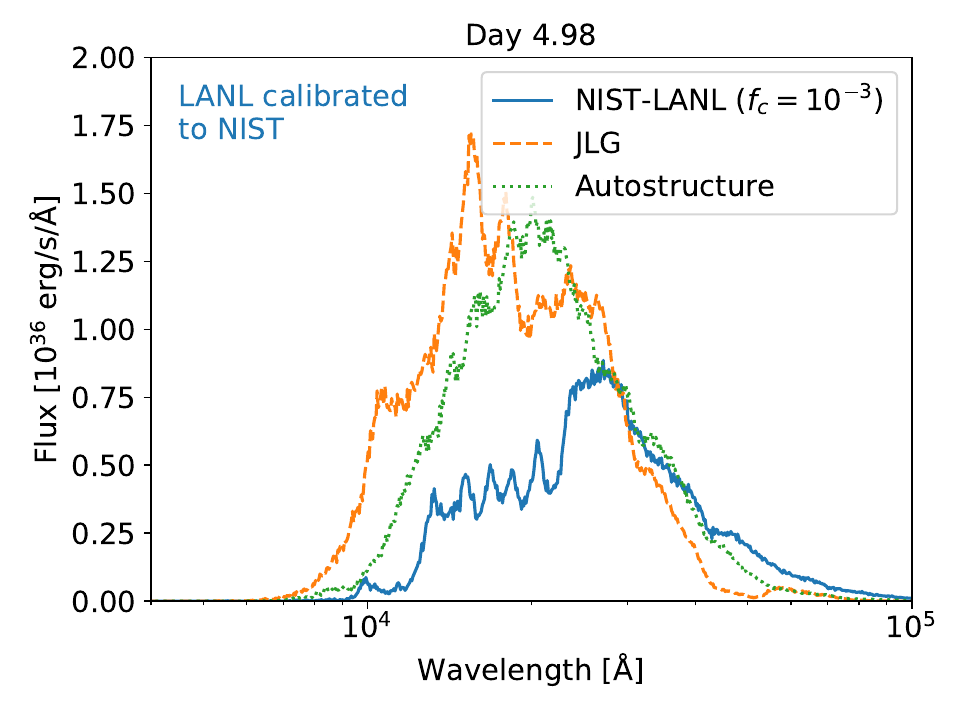} \\
    \includegraphics[width=0.49\linewidth]{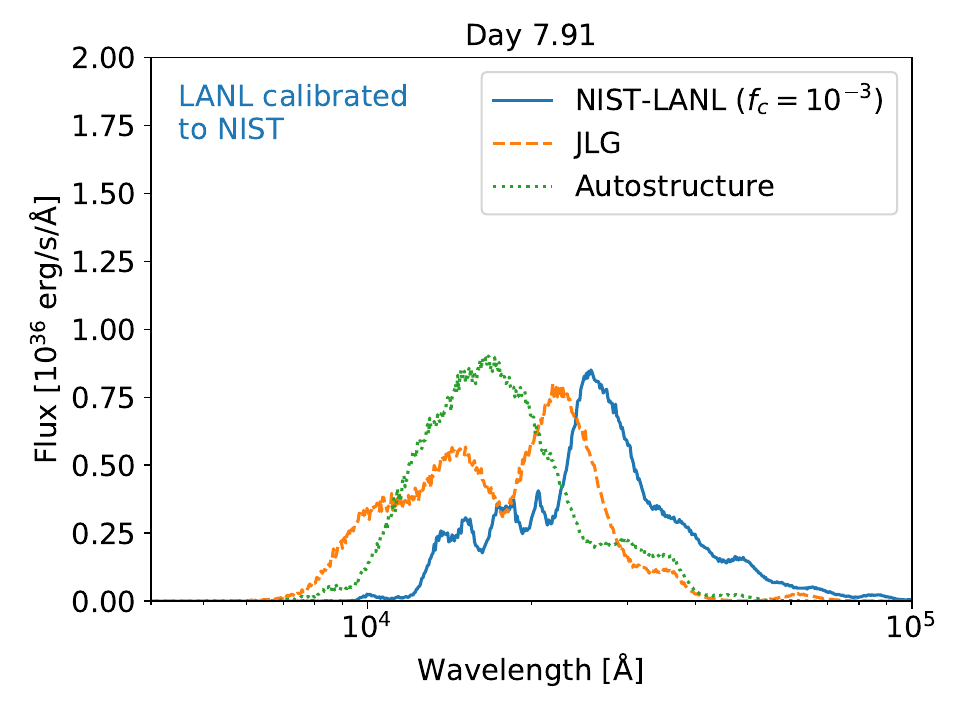}
    \includegraphics[width=0.49\linewidth]{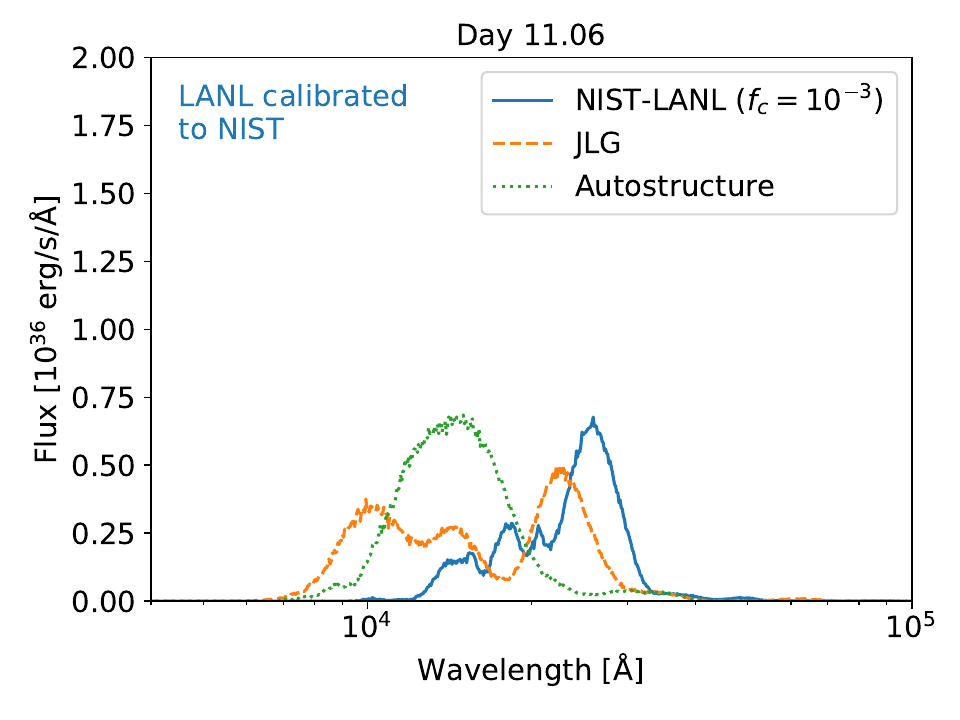}
    \caption{Same as Figure~\ref{fig2:lcspec}, but with LANL data calibrated to the NIST database
    (Section~\ref{sec:lanlsapc}).
    At days 2 and 5 the specta are dim compared to those of the original LANL data set.
    Similar to the NN-LANL data set in Figure~\ref{fig4:lcspec}, where the neutral stage has been
    effectively removed, we observe that the calibration significantly effects mid-IR line
    structure associated with Nd I.}
    \label{fig5:lcspec}
\end{figure}

Finally, we compare the NIST-calibrated version of the LANL ($f_c = 10^{-3}$) data set, which we label
NIST-LANL ($f_c=10^{-3}$).
The method of calibration is described in Section~\ref{sec:callanl}.
Figure~\ref{fig5:lcspec} again has the same data (spectra at days 2, 5, 8 and 11) as Figure~\ref{fig2:lcspec},
but with the results for NIST-LANL data replacing those of the LANL data.
Similar to the NN-LANL test, we see that the mid-IR line structure is significantly impacted by the
calibration, specifically the calibration for Nd I.
The calibration for line data of Nd II-IV has insignificant effect at days 8 and 11, due to the dominance
of the Nd I population by those times.
Unlike the NN-LANL test, we see the early time spectrum is rendered somewhat dimmer across much of the
wavelength range contributing to the bulk spectrum.

\begin{figure}[!h]
    \centering
    \includegraphics[width=0.49\linewidth]{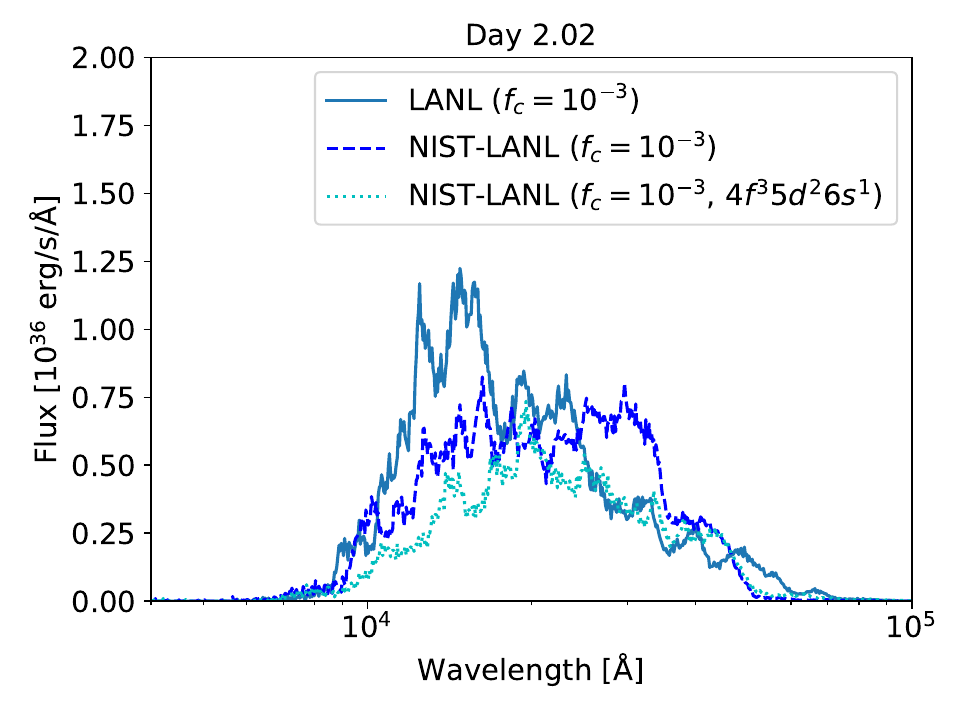}
    \includegraphics[width=0.49\linewidth]{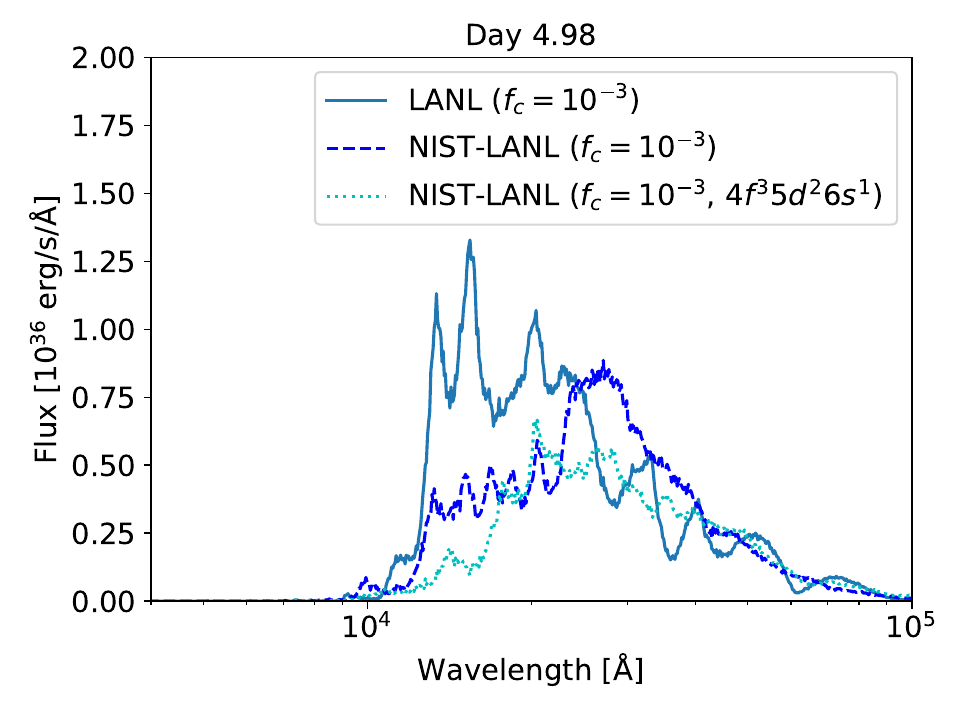} \\
    \includegraphics[width=0.49\linewidth]{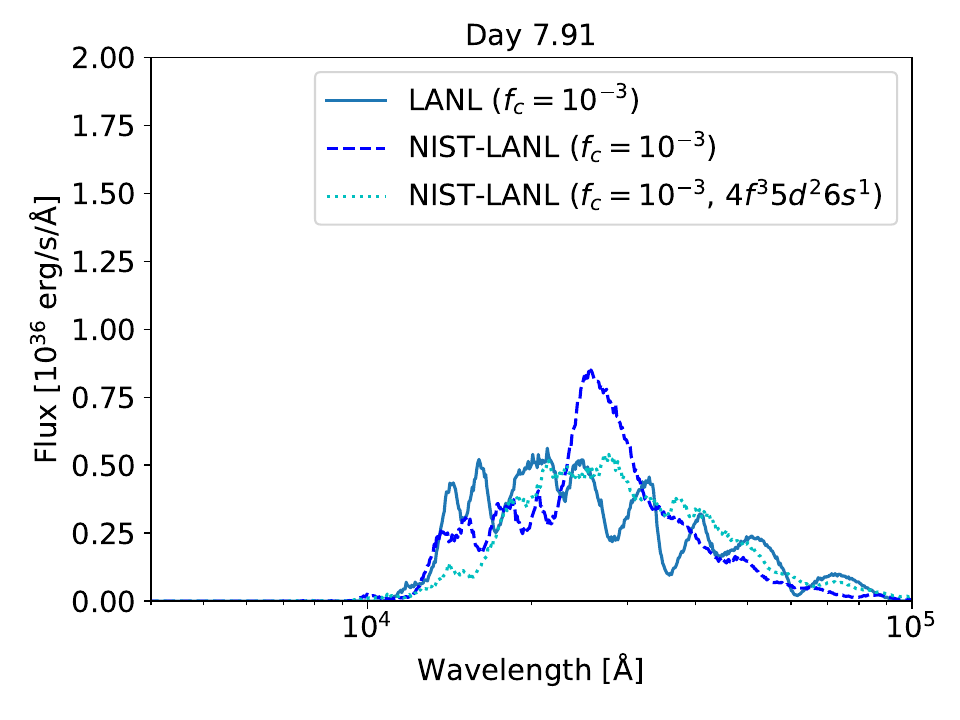}
    \includegraphics[width=0.49\linewidth]{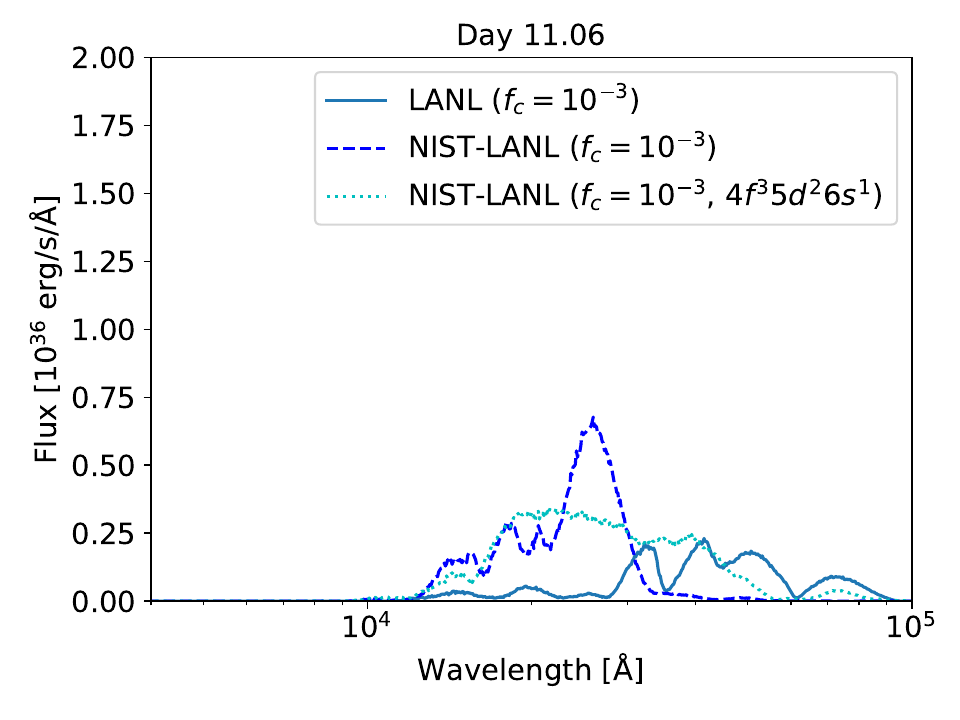}
    \caption{Spectra versus wavelength at day 2 (upper left), day 5 (upper right), day 8 (lower left)
    and day 11 (lower right), for uncalibrated LANL (solid), calibrated NIST-LANL (dashed) data sets, and NIST-LANL data with $4f^35d^26s^1$ added to Nd I (dotted).
    The addition of the $4f^35d^26s^1$ configuration to Nd I significantly affects the spectrum at
    each time shown.}
    \label{fig6:lcspec}
\end{figure}

An indicated in Section~\ref{sec:callanl}, we test a second calibration where we include the 
$4f^35d^26s^1$ electron configuration in Nd I.
Figure~\ref{fig6:lcspec} has spectra at days 2, 5, 8 and 11 for: the uncalibrated LANL data set,
the NIST-LANL data without $4f^35d^26s^1$ in Nd I (dashed), and the NIST-LANL with $4f^35d^26s^1$
in Nd I (dotted).
The addition of the $4f^35d^26s^1$ configuration to Nd I significantly affects the spectrum at each
time shown (though at day 1, when Nd I is a small fraction of the ion population, the effect of the
configuration is small).

\section{Conclusions}
\label{sec:conc}

We simulate a simple pure Nd 1D KN model with three data sets originating from three different
codes: the LANL suite of atomic physics codes~\citep{fontes2015b}, \texttt{HULLAC}~\citep{barshalom2001,tanaka2020}
and \texttt{Autostructure}~\citep{badnell2011,badnell2012,badnell2016}.
Our findings may be summarized as follows. %\todo{Write conclusions last.}
\begin{itemize}
    \item Consistent with previous findings, different opacity data sets can have a significant impact on KN light curves and spectra. These differences may arise from differences in atomic configurations used and in the atomic physics codes themselves, especially in light of the challenges presented by calculations for the lanthanides and actinides.
    
    \item Over the time span simulated, the ionization fractions are in good agreement among the three atomic datasets, but a divergence in ejecta temperature coincides with growth in the Nd I fraction at outer ejecta radii.
    
    \item We find that the neutral stage, where the three calculations show significant differences, has a strong impact on the agreement between light curves and spectra of different data sets, particularly at late time. The LANL data set has the largest number of levels and lines in the neutral stage, with particularly strong lines in the mid-IR.
    
    \item The impact of the neutral stage can be seen explicitly by removing the neutral stage from the simulation, and only performing the radiative transfer with Nd II, III and IV, albeit in the assumption of LTE. For the simple KN model used here, the LANL data with the neutral stage removed shows better agreement with the other codes in terms of bolometric luminosity, the shape of the $r$- and $z$-band light curves, and the bulk emission and mid-IR structure of the spectra. However, the $r$- and $z$-band emission for the no-neutral LANL data still eventually fall below those of the JLG and \texttt{Autostructure} data sets.
    
    \item Calibrating the LANL data set to curated NIST ASD energies, by replacing energy level values after the atomic structure calculation but before oscillator strength evaluation, significantly impacts the synthesized spectra. The spectra are dimmer at early times (near peak luminosity) and show less mid-IR line structure at late times. We find that the late-time differences are due to the calibration of Nd I, consistent with the finding that Nd I dominates the ion population at these times, under the LTE assumptions in {\tt SuperNu}. Furthermore, we see that the inclusion of the $4f^35d^26s^1$~electron configuration for Nd I significantly impacts the spectral structure from day 2 through day 11.
    %\todo{Have Fontes vet this...}
\end{itemize}

The results presented here underscore the need for spectroscopically accurate lanthanide (and actinide)
atomic data in order to better interpret observed KNe. The significant effect of our NIST ASD calibration on KN spectra might appear to be in conflict with the
results of~\cite{flors2023}, which used calibrated energies for singly and doubly ionized Nd, but there
are several important distinctions between the present study and that earlier work. First, we focused on the 
effect of calibration on spectra, which probe a range of thermodynamic conditions and time-dependent emission wavelengths, unlike individual opacities. Second, we also included a calibration of the level energies
for Nd I, which has the most significant impact on our late-time LTE spectra. Finally, our calibration
technique is distinct in that we replace specific energy level values, rather than use an approach that
matches configuration-average energies with the corresponding results that are deduced from experimental level energies.
\bfedit{
In the future, we will compute and calibrate opacities for other elements following similar procedures and make the data available on the NIST-LANL Opacity Database~\citep{ralchenko2020}.
}

Notwithstanding the possibility of suppressed neutral stages for lanthanides and actinides at late time,
accurate atomic levels and lines for neutral stages, in particular, may improve the fidelity of
late-time KN spectra~\citep{hotokezaka2021,pognan2022a,pognan2022b,brethauer2025}.
However, apart from calibration and non-LTE effects, other uncertainties in KN simulations should be systematically
evaluated in relation to the effects of atomic data. One potentially fruitful avenue for exploration would be the composition~\citep{even2020,pognan2025}
and morphology, which in particular has seen much recent attention in 2D and 3D
(see, for example,~\citealt{heinzel2021,korobkin2021,bulla2023,collins2024,fryer2024,kawaguchi2024}, and
references therein).
These multitude uncertainties of course affect the thermodynamic state of the KN ejecta throughout the time scale
of observable evolution, and may complicate determination of which atomic data sets best fit an observation. As we begin to observe a population of kilonovae, it will be crucial to understand the distinct effects of and interplay between these uncertainties.

\section*{Acknowledgement}

We thank D.~Kasen for preliminary conversations concerning \texttt{Autostructure} line lists.
This work was supported by the U.S. Department of Energy through the Los Alamos National Laboratory.
Los Alamos National Laboratory is operated by Triad National Security, LLC, for the National Nuclear Security Administration of the U.S. Department of Energy (contract No. 89233218CNA000001).
The research presented in this article was supported by the Laboratory Directed Research and Development program of Los Alamos National Laboratory under project No.~20240170ER.
This research used resources provided by the Los Alamos National Laboratory Institutional Computing Program, which is supported by the U.S. Department of Energy National Nuclear Security Administration under contract No. 89233218CNA000001. N.V.\ acknowledges funding from the Natural Sciences and Engineering Research Council of Canada (NSERC) Canada Graduate Scholarship - Doctoral (CGS-D) and Postdoctoral Fellowship (PDF).
\bfedit{M.~R. acknowledges support from the Information Science \& Technology Institute (ISTI) at LANL as an ISTI Postdoctoral Fellow and acknowledges PECASE award funds.}

\bibliography{refs}{}
\bibliographystyle{aasjournal}

\end{document}